\documentclass[11pt,a4paper]{article}
\pdfoutput=1

\usepackage{jheppub}
\usepackage{graphicx}
\usepackage{epsfig}
\usepackage{subfig}
\usepackage{url}
\usepackage{dsfont}
\usepackage{amssymb}
\usepackage{slashed}
\usepackage{booktabs,multirow,tabularx}
\usepackage{braket}
\usepackage{bm}
\usepackage{float} 
\usepackage{subfig} 
\usepackage{hyperref}
\usepackage{comment}
\allowdisplaybreaks

\DeclareMathAlphabet{\pazocal}{OMS}{zplm}{m}{n}
\newcommand\sss{}
\newcommand\mydot{\!\cdot\!}

\newcommand\ep{\epsilon}

\def\p0{{\bigl.^3\hspace{-1mm}P^{[8]}_0}}

\def\to{\rightarrow}

\def\bqa{\begin{eqnarray}}
\def\eqa{\end{eqnarray}}
\def\bc{\begin{center}}
\def\bc{\end{center}}

\newcommand\allproc{{\cal R}}

\newcommand\allprocn{\allproc_{n}}

\newcommand\nheavy{n_{\sss H}}

\newcommand\avg{{\cal N}}

\newcommand\ident{{\cal I}}
\newcommand\amp{{\cal A}}

\newcommand\ampnt{\amp^{(n,0)}}
\newcommand\ampnpot{\amp^{(n+1,0)}}

\def\remove#1#2{#1\hspace{-#2truecm}\backslash}

\newcommand\Qop{\vec{Q}}

\newcommand\ampsq{{\cal M}}

\newcommand\ampsqnt{\ampsq^{(n,0)}}

\newcommand\ampsqnpot{\ampsq^{(n+1,0)}}

\newcommand\Ione{\ident_1}
\newcommand\Itwo{\ident_2}

\def\be{\begin{equation}}
\def\ee{\end{equation}}
\def\bea{\begin{eqnarray}}
\def\eea{\end{eqnarray}}

\def\msbar{\overline{\mathrm{MS}}}

\newcommand{\boldirrep}{\mathbf}
\newcommand{\irrepbase}[1]{\ensuremath{\boldirrep{#1}}}

\newlength{\irrepwidth}
\newlength{\irrepbarthickness}
\newlength{\irrepbarheight}
\newcommand{\irrepbarbase}[1]{%
    \settoheight{\irrepbarheight}{\irrepbase{#1}}%
    \settowidth{\irrepwidth}{\irrepbase{#1}}%
    \makebox[0pt][l]{\irrepbase{#1}}%
    \rule[1.2\irrepbarheight]{\irrepwidth}{\irrepbarthickness}%
}

\def\primes#1#2{\count0=#1 \loop \ifnum\count0>0 \advance\count0 by -1 #2\repeat}
\newcommand{\irrep}[2][0]{\ensuremath{\irrepbase{#2}^{\primes{#1}{\prime}}}}
\newcommand{\irrepbar}[2][0]{\ensuremath{\irrepbarbase{#2}^{\primes{#1}{\prime}}}}

\newcommand{\harpoon}{\overset{\rightharpoonup}}

\usepackage{tikz}
\usetikzlibrary{positioning,arrows}
\usetikzlibrary{decorations.pathmorphing}
\usetikzlibrary{decorations.markings}
\usetikzlibrary{shapes.geometric}
\usepackage{endnotes}
\tikzset{
    vector/.style={decorate, decoration={snake}, draw},
    provector/.style={decorate, decoration={snake,amplitude=2.5pt}, draw},
    antivector/.style={decorate, decoration={snake,amplitude=-2.5pt}, draw},
    fermion/.style={draw=black,
      postaction={decorate},decoration={markings,mark=at position .55
        with {\arrow[draw=black]{>}}}},
    fermionbar/.style={draw=black, postaction={decorate},
                       decoration={markings,mark=at position .55 with {\arrow[draw=black]{<}}}},
    fermionnoarrow/.style={draw=black},
    gluon/.style={decorate, draw=black,decoration={coil,amplitude=4pt, segment length=6pt}},
    scalar/.style={dashed,draw=black,
      postaction={decorate},decoration={markings,mark=at position .55
        with {\arrow[draw=black]{>}}}},
    scalarbar/.style={dashed,draw=black,
      postaction={decorate},decoration={markings,mark=at position .55
        with {\arrow[draw=black]{<}}}},
    scalarnoarrow/.style={dashed,draw=black},
    electron/.style={draw=black,
      postaction={decorate},decoration={markings,mark=at position .55
        with {\arrow[draw=black]{>}}}},
    bigvector/.style={decorate, decoration={snake,amplitude=4pt}, draw},
}

\newcommand{\backslashover}[1]{%
  \begin{tikzpicture}[baseline=(X.base)]
    \node (X) at (0,0) {\(#1\)};
    \draw (-0.32,0.05) -- (0.32,-0.05);  
  \end{tikzpicture}
}

\normalbaselines

\title{One-loop transverse-momentum-dependent soft function at higher orders in the dimensional regulator}

\author{Hua-Sheng Shao and Guoxing Wang}
\emailAdd{huasheng.shao@lpthe.jussieu.fr,wangguoxing2015@pku.edu.cn}
\affiliation{Laboratoire de Physique Th\'eorique et Hautes Energies (LPTHE), UMR 7589, Sorbonne Universit\'e et CNRS, 4 place Jussieu, 75252 Paris Cedex 05, France
}

\preprint{}

\abstract{The transverse-momentum-dependent (TMD) soft function for a generic hadroproduction process involving massive colored particles is analytically calculated at the one-loop level, extended to higher orders in the dimensional regulator $\epsilon$. We present both the azimuthal-angle-averaged and azimuthal-angle-dependent soft functions in impact-parameter space, making them suitable for small $q_T$ resummation calculations. Their analytic expressions are provided in terms of multiple polylogarithms. Our results offer essential ingredients for a complete higher-order perturbative calculation of the TMD soft function.}

\begin{document}
\maketitle
\flushbottom

\section{Introduction}

Soft functions are indispensable in establishing factorization formalisms for cross sections and in perturbative resummation calculations. In quantum chromodynamics (QCD), they encode the dynamics of soft parton (i.e., massless quarks, antiquarks, and gluons) exchanges between colored external legs in the $S$-matrix, leading to non-trivial color correlations. Soft functions are universal in the sense that they are fully determined by the quantum numbers and kinematics of external colored particles. Among the process-independent components of the factorization theorem, they are by far the most intricate. For hadroproduction processes involving only color-singlet particles in the final state, soft functions remain relatively simple, and analytic expressions are usually known. However, when multiple colored partons are present, their structure becomes significantly more complex. 

At facilities such as the Large Hadron Collider (LHC) at CERN and the planned Electron-Ion Collider (EIC) at BNL in the US, understanding small-transverse-momentum ($q_T$) physics with high precision is particularly important. It encodes key information about the multi-dimensional internal structure and spin of hadrons and could even provide unique opportunities to uncover elusive Beyond the Standard Model (BSM) signals. At small $q_T$ relative to the hard scale $M$, fixed-order perturbative computations of cross sections suffer from large theoretical uncertainties due to presence of large logarithms, $\log{\left(M^2/q_T^2\right)}$, which must be resummed to all orders in the strong coupling constant $\alpha_s$. Small-$q_T$ resummation formalisms, which can be formulated in various ways, generally require knowledge of the transverse-momentum-dependent (TMD) soft function, particularly in processes involving colored particles in the final state.

Generally speaking, the TMD soft function depends on both the magnitude $q_T$ and the azimuthal angle $\phi_q$ of the two-dimensional transverse momentum $\harpoon{q}_T$ of the final system. If one is primarily interested in a $q_T$-dependent cross section integrated over the azimuthal angle, the azimuthally averaged TMD soft function is often sufficient for many applications. On the other hand, the full azimuth-dependent soft function is essential for studying azimuthal (de-)correlations of final-state particles in the small-$q_T$ region. Beyond small-$q_T$ resummation calculations, the TMD soft function also plays a significant role in the $q_T$ slicing approach~\cite{Catani:2007vq}, which is used to handle infrared (IR) divergences in fixed-order perturbative calculations beyond leading order (LO). Originally designed for processes with colorless final states, this approach has been successfully extended to include massive colored particles at next-to-next-to-leading order (NNLO)~\cite{Bonciani:2015sha}. A generalization to jet processes has recently been proposed at next-to-leading order (NLO)~\cite{Fu:2024fgj}. In most applications of the slicing method, only the azimuthally averaged soft function is required. Overall, both azimuthal-angle-averaged and azimuthal-angle-dependent soft functions are valuable universal quantities.

Despite the soft function having a well-defined operator formulation in quantum field theory, it is not a physical quantity. This has at least two important consequences. First, evaluating the soft function at one-loop order and beyond introduces IR singularities, requiring regularization. In this paper, we use dimensional regularization with spacetime dimension $d=4-2\epsilon$, where $\epsilon$ is the dimensional regulator. Second, due to these divergences, the finite remainder of the soft function is inherently scheme-dependent in resummation frameworks. However, physical observables, such as cross sections (which involve a combination of the soft function and other quantities), remain formally independent of the chosen scheme.
Many calculations of the TMD soft function, both numerical and analytical, have been performed in the literature. Relevant to our discussions, we note that at the one-loop level, the azimuthally-averaged and azimuthal-angle-dependent soft functions were first computed in refs.~\cite{Li:2013mia} and \cite{Catani:2014qha}, respectively, up to $\mathcal{O}(\epsilon^0)$ in the Laurent series expansion of the dimensional regulator, for the production of a back-to-back heavy quark pair. The kinematic constraint of back-to-back motion in the transverse plane simplifies the calculations, and the soft functions are expressed in terms of multiple polylogarithms. Subsequently, for the same process, the $\mathcal{O}(\epsilon)$ one-loop terms and the two-loop contributions up to $\mathcal{O}(\epsilon^0)$ were computed in refs.~\cite{Angeles-Martinez:2018mqh,Catani:2023tby} for the azimuthally-averaged soft function. While the one-loop $\mathcal{O}(\epsilon)$ terms are known analytically in terms of multiple polylogarithmic functions, the two-loop results have been provided only numerically in a grid format, as some integrals require numerical evaluation. The back-to-back constraint was lifted in refs.~\cite{Catani:2021cbl,Ju:2022wia} for the azimuthal-angle-dependent soft function at one loop, up to $\mathcal{O}(\epsilon^0)$. However, some remaining one-fold integrals still need to be evaluated numerically. The one-loop results obtained there are, in principle, applicable to any process without jets in the final state. The findings in ref.~\cite{Devoto:2024nhl} suggest that the two-loop azimuthal-angle-averaged soft function for hadroproduction of a heavy quark pair in association with any colorless particles has been (at least numerically) computed, as the NNLO $q_T$ slicing approach was employed. However, the soft function has not yet been published. Recently, the two-loop fully differential soft function for $e^+e^-\to Q\bar{Q}V$, where $V$ is any color-singlet system, was presented in ref.~\cite{Liu:2024hfa}. The paper provides the azimuthal-angle-dependent TMD soft function for this process at one-loop $\mathcal{O}(\epsilon)$ and two-loop $\mathcal{O}(\epsilon^0)$ in analytic form, expressed in terms of multiple polylogarithms. Unlike traditional approaches to computing the TMD soft function, the authors first convert the eikonal integrals into standard Feynman integrals and then apply the established multi-loop Feynman integral techniques. This approach is compelling due to its potential for generalization to more complex processes and higher-loop orders.

In this paper, we analytically compute the one-loop azimuthal-angle-averaged and azimuthal-angle-dependent TMD soft functions for arbitrary processes involving massive colored and colorless particles, extending known results to higher orders in $\epsilon$. This calculation serves as an interesting exercise and provides essential ingredients for a complete N$^{k+1}$LO computation, where the one-loop TMD soft function at $\mathcal{O}(\epsilon^k)$ with $k\geq 0$ is required. Our results not only extend the known $\mathcal{O}(\epsilon^0)$ expression in ref.~\cite{Catani:2021cbl} to a fully analytic form but also provide new representations in terms of multiple polylogarithms at $\mathcal{O}(\epsilon^k)$ for $k\geq 1$. Our approach is systematic and, in principle, allows for the computation of results at any order in $\epsilon$. However, in this work, we explicitly present results up to $\mathcal{O}(\epsilon^3)$.

The paper is organized as follows. In section \ref{sec:factorization}, we discuss the factorization formula in the small-$q_T$ limit and introduce light-cone coordinates and power counting. In section \ref{sec:softfunction}, we present the analytic results for the one-loop TMD soft function at higher orders in $\epsilon$ and detail our computational approach. Finally, we summarize our findings in section \ref{sec:summary}.

\section{Factorization formula\label{sec:factorization}}

Following the notation of refs.~\cite{Frederix:2009yq,Alwall:2014hca,Frederix:2018nkq,AH:2024ueu,Shao:2025fwd}, we consider a generic $2\to n$ scattering process at a hadron collider, with the partonic Born subprocess denoted as $\Ione\Itwo\to \ident_3\cdots \ident_{n+2}$, where $\ident_i$ represents the identity of the $i$th particle. This subprocess can be expressed as an ordered list $r=(\Ione,\Itwo,\cdots,\ident_{n+2})$, and the four-momentum of each external particle $\ident_i$ is denoted as $k_i$. In the absence of real radiation, momentum conservation reads
\begin{equation}
k_1+k_2=\sum_{i=3}^{n+2}{k_i}\,.
\end{equation}
The set of all possible subprocesses defines a space $\allprocn$, with $r\in \allprocn$. In this paper, we exclude Born processes involving massless colored particles in the final state, meaning that jets are not considered. Without loss of generality, for any given $r\in\allprocn$, we reorder the final-state particles such that strongly interacting massive particles are indexed by $3\leq j\leq \nheavy+2$, where $\nheavy$ ($0\leq\nheavy\leq n$) denotes the number of massive colored partons in the final state. The remaining particles, indexed from $\nheavy+3$ to $n+2$, are colorless. In the Standard Model (SM), the only possible initial-state partons in this configuration are gluons and quark-antiquark pairs, as we explicitly exclude (anti-)quark-gluon channels, such as in the process $bg\to tW^-$.

When considering additional real parton radiation, the final-state Born system acquires a nonzero transverse momentum, i.e.,
\begin{equation}
\sum_{i=3}^{n+2}{\harpoon{k}_{i,T}}=\harpoon{q}_T,
\end{equation}
where $\harpoon{k}_{i,T}$ is the $(d-2)$-dimensional (or two-dimensional when setting $d=4$) transverse momentum vector of particle $\ident_i$. Throughout this paper, we denote $\harpoon{x}_{T}$ as the $(d-2)$-dimensional transverse vector of any $d$-dimensional quantity $x^\mu$. In the small $q_T\equiv |\harpoon{q}_T|$ region, the factorization formula~\cite{Becher:2010tm,Catani:2014qha,Becher:2012yn,Zhu:2012ts,Li:2013mia} for small-$q_T$ resummation, based on soft-collinear effective field theory (SCET)~\cite{Bauer:2000ew,Bauer:2000yr,Bauer:2001ct,Bauer:2001yt}, is given by:
\begin{eqnarray}
\frac{d\sigma}{d^2\harpoon{q}_TdYdM^2d\phi_n}&=&\frac{1}{S}\int{dx_1dx_2 \delta\left(x_1x_2-\frac{M^2}{S}\right)\delta\left(\frac{1}{2}\log{\left(\frac{x_1}{x_2}\right)}-Y\right)}\nonumber\\
&&\times\int{\frac{d^2\harpoon{b}_{T}}{\left(2\pi\right)^2}e^{i\harpoon{q}_T\cdot \harpoon{b}_T}W(x_1,x_2,M,\harpoon{b}_T)}\,,\label{eq:qtresumgeneralinb}
\end{eqnarray}
where $S$ is the squared center-of-mass energy of the two colliding hadrons $N_1$ and $N_2$, and $Y$~\footnote{$Y$ is defined in the center-of-mass frame of the two hadrons $N_1$ and $N_2$.} and $M$ are the rapidity and invariant mass of the Born-level final state, respectively. The variables $x_1$ and $x_2$ are the longitudinal momentum fractions of the two initial-state partons $\Ione$ and $\Itwo$ in the Born subprocess. The measure $d\phi_n$ represents the phase-space element of the Born-like $n$-body system. The impact-parameter-dependent function $W()$ in eq.~\eqref{eq:qtresumgeneralinb} can be expressed as
\begin{eqnarray}
W(x_1,x_2,M,\harpoon{b}_T)&=&\left(\frac{b_T^2 M^2}{b_0^2}\right)^{-F_{gg}(b_T^2,\mu)}4B_{g/N_1}^{\alpha\rho}(x_1,\harpoon{b}_T,\mu)B_{g/N_2}^{\beta\sigma}(x_2,\harpoon{b}_T,\mu)\nonumber\\
&&\times{\rm Tr}_c\left[\bm{H}_{gg,\alpha\beta\rho\sigma}(M,\mu)\bm{S}_{\perp,gg}(\harpoon{b}_T,\mu)\right]\nonumber\\
&&+\left(\frac{b_T^2 M^2}{b_0^2}\right)^{-F_{qq}(b_T^2,\mu)}\sum_{i,j=1}^{n_q}{B_{q_i/N_1}(x_1,b_T,\mu)B_{\bar{q}_j/N_2}(x_2,b_T,\mu)}\nonumber\\
&&\times{\rm Tr}_c\left[\bm{H}_{q_i\bar{q}_j}(M,\mu)\bm{S}_{\perp,q_i\bar{q}_j}(\harpoon{b}_T,\mu)\right]\nonumber\\
&&+\left(\frac{b_T^2 M^2}{b_0^2}\right)^{-F_{qq}(b_T^2,\mu)}\sum_{i,j=1}^{n_q}{B_{\bar{q}_i/N_1}(x_1,b_T,\mu)B_{q_j/N_2}(x_2,b_T,\mu)}\nonumber\\
&&\times{\rm Tr}_c\left[\bm{H}_{\bar{q}_iq_j}(M,\mu)\bm{S}_{\perp,\bar{q}_iq_j}(\harpoon{b}_T,\mu)\right],\label{eq:Wgeneraldef}
\end{eqnarray}
where ${\rm Tr}_c$ denotes the trace over color space, $b_0=2e^{-\gamma_E}$ with $\gamma_E$ being the Euler-Mascheroni constant, and $b_T=|\harpoon{b}_T|$. The sum runs over the indices $i,j$, which range from $1$ to $n_q$, the number of massless quark flavors. The exponential factors involving $F_{\ident\ident}()$ in eq.~\eqref{eq:Wgeneraldef} arise due to the collinear anomaly~\cite{Becher:2010tm}, which prevents a na\"ive factorization of
the cross section at small $q_T$. The function $B_{\ident/N}()$ is the TMD beam function, while $\bm{H}_{\Ione\Itwo}()$ and $\bm{S}_{\perp,\Ione\Itwo}()$ are the hard and soft functions, respectively. The hard and soft functions are generally matrices in the color space spanned by a complete color basis of the underlying Born subprocess $r\in\allprocn$. Let us assume a complete and orthogonal color basis as $|C_l\rangle$ with $l=1,\ldots,n_c$, satisfying 
\begin{equation}
\langle C_{l^\prime}|C_l\rangle=\delta_{l^\prime, l}\langle C_l|C_l\rangle,\quad \bm{1}=\sum_{l=1}^{n_c}{\frac{|C_l\rangle \langle C_l|}{\langle C_l|C_l\rangle}}\,,
\end{equation}
where $\delta_{l,l^\prime}$ is the Kronecker delta function. The choice of color basis is not unique and depends on the partonic subprocess $r$. The trace operator ${\rm Tr}_c$ in eq.~\eqref{eq:Wgeneraldef} ensures that the combination of the hard and soft functions is independent of the specific choice of basis, and consequently, so is $W()$. Although each individual function appearing in eq.~\eqref{eq:Wgeneraldef} depends on the renormalization scale $\mu$, the function $W()$ itself is renormalization-group (RG) invariant. The large logarithms appearing in its perturbative expansion can be resummed using the RG equations.

For a given subprocess $r\in\allprocn$, let us denote the $\ell$-loop, $n$-body ultraviolet (UV) renormalized amplitude as $\mathcal{A}^{(n,\ell)}(r)$ for the quark-antiquark channel ($\Ione,\Itwo=q_i,\bar{q}_j$ or $\bar{q}_i,q_j$), and as $\varepsilon^\mu(k_1)\varepsilon^\nu(k_2)\mathcal{A}^{(n,\ell)}_{\mu\nu}(r)$ for the gluon-fusion channel ($\Ione=\Itwo=g$), where $\varepsilon^\mu(k)$ represents the external polarization vector of a gluon with momentum $k$. The $(l,l^\prime)$ element of the $\ell$-loop hard function in the color space is given by
\begin{equation}
\begin{aligned}
\bm{H}^{(\ell)}_{q\bar{q},ll^\prime}(M,\mu)=&\frac{1}{\avg\omega(\Ione)\omega(\Itwo)}\frac{1}{2M^2}\sum_{i=0}^{\ell}{\frac{1}{1+\delta_{i,\ell-i}}\times}\\
&\frac{\langle C_l|\mathcal{A}^{(n,i)}(r)\rangle \langle \mathcal{A}^{(n,\ell-i)}(r)|C_{l^\prime}\rangle+\langle C_l|\mathcal{A}^{(n,\ell-i)}(r)\rangle \langle \mathcal{A}^{(n,i)}(r)|C_{l^\prime}\rangle}{\langle C_{l^\prime}|C_{l^\prime}\rangle\langle C_{l}|C_{l}\rangle}\,\label{eq:HF4qq}
\end{aligned}
\end{equation}
for the quark-antiquark channel, and
\begin{equation}
\begin{aligned}
\bm{H}^{(\ell)}_{gg,\alpha\beta\rho\sigma,ll^\prime}(M,\mu)=&\frac{1}{\avg\omega(\Ione)\omega(\Itwo)}\frac{1}{2M^2}\sum_{i=0}^{\ell}{\frac{1}{1+\delta_{i,\ell-i}}\times}\\
&\frac{\langle C_l|\mathcal{A}^{(n,i)}_{\alpha\beta}(r)\rangle \langle \mathcal{A}^{(n,\ell-i)}_{\rho\sigma}(r)|C_{l^\prime}\rangle+\langle C_l|\mathcal{A}^{(n,\ell-i)}_{\alpha\beta}(r)\rangle \langle \mathcal{A}^{(n,i)}_{\rho\sigma}(r)|C_{l^\prime}\rangle}{\langle C_{l^\prime}|C_{l^\prime}\rangle\langle C_{l}|C_{l}\rangle}\,\label{eq:HF4gg}
\end{aligned}
\end{equation}
for the gluon-fusion channel. Here, $\avg$ accounts for the final-state symmetry factor, and $\omega(\ident)$ represents the spin and color degrees of freedom of particle $\ident$, given by $\omega(q)=\omega(\bar{q})=2N_c=6$ and $\omega(g)=2(1-\epsilon)D_A=16(1-\epsilon)$ in $d=4-2\epsilon$ within conventional dimensional regularization (CDR). The spin summation of external particles in eqs.\eqref{eq:HF4qq} and \eqref{eq:HF4gg} has been implicitly performed at the amplitude-squared level.

The TMD soft function can be expanded in powers of the small strong coupling constant $\alpha_s$:
\begin{equation}
\bm{S}_{\perp,\Ione\Itwo}(\harpoon{b}_T,\mu)=\sum_{\ell=0}^{+\infty}{\bm{S}_{\perp,\Ione\Itwo}^{(\ell)}(\harpoon{b}_T,\mu)}\,.\label{eq:softfunctionexpandinas}
\end{equation}
Note that in eq.~\eqref{eq:softfunctionexpandinas}, we have not explicitly factored out powers of $\alpha_s$. Instead, each term $\bm{S}_{\perp,\Ione\Itwo}^{(\ell)}(\harpoon{b}_T,\mu)$ implicitly includes an overall factor of $\alpha_s^\ell$. At LO, the TMD soft function is independent of kinematic variables:
\begin{equation}
\bm{S}^{(0)}_{\perp,\Ione\Itwo,l^\prime l}(\harpoon{b}_T,\mu)=\langle C_{l^\prime}|C_l\rangle=\delta_{l^\prime,l}\langle C_{l}|C_l\rangle\,.\label{eq:SFLO}
\end{equation}

At transverse separation $b_T\ll\Lambda^{-1}_{\mathrm{QCD}}$ (where $\Lambda_{\mathrm{QCD}}$ is the intrinsic QCD scale), the TMD beam function can be related to the standard collinear parton distribution function (PDF) via the operator-product expansion (OPE)~\cite{Collins:1981uk,Collins:1981uw,Collins:1984kg}
\begin{eqnarray}
B_{g/N}^{\alpha\beta}(x,\harpoon{b}_T,\mu)&=&
\sum_{\ident}{\int_{x}^1{\frac{dz}{z}\Bigg[\frac{g^{\alpha\beta}_{\perp}}{d-2}I_{g\ident}\left(z,L_\perp\right)+\left(\frac{g^{\alpha\beta}_{\perp}}{d-2}+\frac{\harpoon{b}_T^\alpha \harpoon{b}_T^\beta}{b_T^2}\right)I_{g\ident}^\prime\left(z,L_\perp\right)\Bigg]f^{(N)}_{\ident}\left(\frac{x}{z},\mu^2\right)}}\nonumber\\
&&+\mathcal{O}\left(b_T^2\Lambda_{{\rm QCD}}^2\right)\,,\nonumber\\
B_{q/N}(x,b_T,\mu)&=&\sum_{\ident}{\int_{x}^{1}{\frac{dz}{z}I_{q\ident}\left(z,L_\perp\right)f^{(N)}_{\ident}\left(\frac{x}{z},\mu^2\right)}}+\mathcal{O}\left(b_T^2\Lambda_{{\rm QCD}}^2\right)\,,
\end{eqnarray}
where $f^{(N)}_{\ident}()$ is the $\msbar$ PDF of the hadron $N$ for the parton $\ident$, and the metric is defined as $g_\perp^{\alpha\beta}={\rm diag}\left(0,-\harpoon{1},0\right)$. The logarithmic dependence on the transverse separation is encoded in
\begin{equation}
L_\perp = \log{\left(\frac{b_T^2\mu^2}{b_0^2}\right)}\,.
\end{equation}
The matching coefficients $I_{\ident_1\ident_2}()$ and $I_{g\ident}^\prime()$ describe the perturbative transition from the collinear PDFs to the TMD beam functions. At LO, they take the form
\begin{eqnarray}
I_{\ident_1\ident_2}\left(z,L_\perp\right)&=&\delta\left(1-z\right)\delta_{\ident_1,\ident_2}+\mathcal{O}\left(\alpha_s\right),\quad I^\prime_{g\ident}\left(z,L_\perp\right)=\mathcal{O}(\alpha_s).\label{eq:TMDkernelsLO}
\end{eqnarray}
These matching coefficients have been computed to higher orders in $\alpha_s$ in the literature. The quark coefficient $I_{q\ident}()$ is known up to next-to-next-to-next-to-leading order (N$^3$LO) at $\mathcal{O}(\alpha_s^3)$~\cite{Luo:2019szz,Ebert:2020yqt}, with expressions at NLO~\cite{Becher:2010tm} and NNLO~\cite{Gehrmann:2012ze,Catani:2012qa,Echevarria:2016scs,Luo:2019hmp} having been derived earlier. For the gluon TMD beam function, both helicity-conserving ($I_{g\ident}()$) and helicity-flip ($I_{g\ident}^\prime()$) components contribute. The helicity-conserving gluon kernel $I_{g\ident}()$ is known at NLO~\cite{Becher:2012qa,Chiu:2012ir,Becher:2012yn}, NNLO~\cite{Catani:2011kr,Gehrmann:2014yya,Echevarria:2016scs,Luo:2019bmw}, and N$^3$LO~\cite{Ebert:2020yqt,Luo:2020epw} in QCD. In contrast, the helicity-flip kernel $I_{g\ident}^\prime()$ has been computed at $\mathcal{O}(\alpha_s)$ in refs.~\cite{Catani:2010pd,Becher:2012yn} and at $\mathcal{O}(\alpha_s^2)$ in refs.~\cite{Gutierrez-Reyes:2019rug,Luo:2019bmw}. Notably, the azimuthal-angle dependence in the beam functions arises only through the helicity-flip Lorentz tensor multiplying $I_{g\ident}^\prime()$ in the gluon case.~\footnote{The existence of spin correlations in gluon-induced processes, such as $gg\to\gamma\gamma$, has been discussed in earlier works (see,e.g., ref.~\cite{Balazs:2007hr}).}

It should be noted that we have taken $d=4$ in the Fourier transform in eq.~\eqref{eq:qtresumgeneralinb}, which is justified because $W()$ in eq.~\eqref{eq:Wgeneraldef} is free of any divergence, whereas the beam, soft, and hard functions contain IR poles. 

In many cases, we are not concerned with the azimuthal-angle distribution of $\harpoon{q}_T$ and are only interested in its magnitude, $q_T=\left|\harpoon{q}_T\right|$. In such situations, the factorization formalism \eqref{eq:qtresumgeneralinb} can be rewritten as
\begin{eqnarray}
\frac{d\sigma}{dq_T^2 dYdM^2d\phi_n}&=&\frac{1}{2S}\int{dx_1dx_2 \delta\left(x_1x_2-\frac{M^2}{S}\right)\delta\left(\frac{1}{2}\log{\frac{x_1}{x_2}}-Y\right)}\nonumber\\
&&\times\int{\frac{d^2\harpoon{b}_T}{(2\pi)^2}2\pi J_0\left(b_T q_T\right)W(x_1,x_2,M,\harpoon{b}_T)}\,,\label{eq:qtresumgeneralinbqT}
\end{eqnarray}
where $J_0()$ is the Bessel function of the first kind $J_n()$ with order $n=0$. Since the beam function $B_{q/N}()$ depends only on $b_T=|\harpoon{b}_T|$ and not on the azimuthal angle of $\harpoon{b}_T$, and in many cases--such as at NLO or next-to-next-to-leading logarithmic (NNLL) accuracies-- the term $I_{g\ident}^\prime()\bm{S}^{(1)}_{\perp,gg}()$ does not contribute, it is often convenient to work with the azimuthal-angle-averaged soft function, which typically has a simpler analytic expression:
\begin{eqnarray}
\bm{S}_{\perp,\Ione\Itwo}(b_T,\mu)&=& \int{\frac{d\Omega_b^{(d-2)}}{\Omega_b^{(d-2)}}\bm{S}_{\perp,\Ione\Itwo}(\harpoon{b}_T,\mu)}\,.\label{eq:Softavgdef}
\end{eqnarray}
where the measure is given by
\begin{equation}
d\Omega_b^{(d-2)}=\prod_{i=1}^{d-3}{d\varphi_i\left(\sin{\varphi_i}\right)^{d-3-i}}\,
\end{equation}
with the corresponding integration result
\begin{equation}
\Omega_b^{(d-2)}=\int{d\Omega_b^{(d-2)}}=\frac{2\pi^{\frac{d-2}{2}}}{\Gamma\left(\frac{d-2}{2}\right)}\,.
\end{equation}
Additionally, we will use the following solid-angle integrals:
\begin{eqnarray}
\int{d\Omega_b^{(d-2)}e^{\pm i\harpoon{b}_T\cdot \harpoon{q}_T}}&=&\left(2\pi\right)^{\frac{d-2}{2}}\left(b_Tq_T\right)^{\frac{4-d}{2}}J_{\frac{d-4}{2}}\left(b_Tq_T\right),\nonumber\\
\int{\frac{d\Omega_b^{(d-2)}}{\Omega_b^{(d-2)}}e^{\pm i\harpoon{b}_T\cdot \harpoon{q}_T}}&=&\left(2\right)^{\frac{d-4}{2}}\Gamma\left(\frac{d-2}{2}\right)\left(b_Tq_T\right)^{\frac{4-d}{2}}J_{\frac{d-4}{2}}\left(b_Tq_T\right).\label{eq:solidangleFT}
\end{eqnarray}

Since it is convenient to work with light-cone coordinates, let us introduce them here. Any $d$-dimensional momentum $k^\mu$ can be decomposed into
\begin{eqnarray}
k^\mu&=&\underbrace{n\cdot k}_{\equiv k_-} \frac{\bar{n}^\mu}{2}+\underbrace{\bar{n}\cdot k}_{\equiv k_{+}}\frac{n^\mu}{2}+\underbrace{k_\perp^\mu}_{=g_\perp^{\mu\nu}k_{\nu}}\equiv k_{-}^\mu+k_{+}^\mu+k_\perp^\mu,
\end{eqnarray}
where the light-like directions of the two beams are defined as 
\begin{eqnarray}
n&=&(1,\underbrace{0,\cdots,0}_{d-2},1)=(1,\harpoon{0},1)\,,\\
\bar{n}&=&(1,\underbrace{0,\cdots,0}_{d-2},-1)=(1,\harpoon{0},-1)\,,
\end{eqnarray}
and $k_{\pm}=k^0\pm k_z$. Using $n^2=\bar{n}^2=0$ and $n\cdot \bar{n}=2$,  we can easily deduce that
\begin{eqnarray}
p\cdot k&=&\frac{p_+k_{-}+p_{-}k_{+}}{2}+p_\perp\cdot k_\perp=p_+\cdot k_{-}+p_{-}\cdot k_{+}+p_\perp\cdot k_\perp\,,
\end{eqnarray}
where
\begin{eqnarray}
p_\perp\cdot k_\perp&=&g^{\mu\nu}_\perp p_{\mu} k_{\nu}=g^{\mu\nu}_\perp p_{\perp,\mu} k_{\perp,\nu}=-\harpoon{p}_T\cdot \harpoon{k}_T\,.
\end{eqnarray}
In light-cone coordinates, the squared momentum is given by
\begin{eqnarray}
k^2&=&k_{+}k_{-}-k_{T}^2\,,
\end{eqnarray}
and the integration measure takes the form
\begin{eqnarray}
d^dk&=&\frac{1}{2}dk_{+}dk_{-}d^{d-2}\harpoon{k}_{T}\,.
\end{eqnarray}
Furthermore, we have the integral
\begin{eqnarray}
\int d^dk\delta^{(+)}(k^2)&=&\frac{1}{2}\int_0^{\infty}dk_{+}dk_{-}\int_{-\infty}^{\infty}d^{d-2}\harpoon{k}_{T}\,,
\label{eq:intmeasure1}
\end{eqnarray}
where $\delta^{(+)}(k^2) = \delta(k^2)
\Theta(k^0)$, with $\Theta()$ being the Heaviside step function.

For a generic scattering process with a hard scale of order $M$, we consider $0\leq \lambda\equiv k_T/M\ll 1$. In this case, the following momentum modes are relevant for small $q_T$ resummation in SCET:
\begin{eqnarray}
\text{hard}\quad &:& \quad (k_-,k_+,\harpoon{k}_T)\sim M(1,1,\harpoon{1})\,,\\
\text{hard-collinear}\quad &:& \quad (k_-,k_+,\harpoon{k}_T)\sim M(\lambda^2,1,\harpoon{\lambda})\,,\\
\text{anti-hard-collinear}\quad &: & \quad (k_-,k_+,\harpoon{k}_T)\sim M(1,\lambda^2,\harpoon{\lambda})\,,\\
\text{soft}\quad &:& \quad (k_-,k_+,\harpoon{k}_T)\sim M(\lambda,\lambda,\harpoon{\lambda})\,.
\end{eqnarray}

In the context of SCET, the full QCD Feynman integrals are approximated following the expansion by regions with the power counting described above. A special type of divergence, called ``rapidity divergences"~\cite{Collins:2003fm}, appears in the intermediate steps of TMD beam and soft function calculations. These rapidity divergences cannot be regularized using dimensional regularization. They arise from the momentum region where the invariant mass $k^2$ is held fixed, but the ratio $k_-/k_+$ or $k_+/k_-$ diverges. In SCET, the separation between soft and collinear modes is arbitrary, which leads to the appearance of rapidity divergences~\cite{Chiu:2012ir}. Therefore, rapidity regulators must be introduced in addition to the dimensional regulator $\epsilon$. In this paper, we choose the analytic rapidity regulator~\cite{Becher:2010tm,Becher:2011dz}, which leads to the following replacement of the phase-space integration measure:
\begin{equation}
\int{\frac{d^dk_g}{(2\pi)^d}2\pi \delta^{(+)}(k_g^2)}\quad \to \quad \int{\frac{d^dk_g}{(2\pi)^d}2\pi \delta^{(+)}(k_g^2)\left(\frac{\nu_1}{n\cdot k_g-i0^+}\right)^{\alpha_1}}\,,\label{eq:analyticrapidityregulation}
\end{equation}
where $\nu_1$ is a dimensionful rapidity regularization scale, and $\alpha_1$ is the rapidity regulator. Note that $n\cdot k_g > 0 $ since $k_g$ has a nonzero transverse component in our case. As a result, the $-i0^+$ prescription in eq.~\eqref{eq:analyticrapidityregulation} can be safely removed. The introduction of the rapidity regulator $\alpha_1$ becomes necessary at intermediate stages. Since rapidity divergences do not appear in the full theory, the cross section should be finite in the limit $\alpha_1\to 0$. For the analytic rapidity regulator, the pole cancellation in $\alpha_1$ is already manifested at level of the soft function,~\footnote{This would no longer be true if one considers a process with two initial-state particles having non-identical Casimir factors, such as in the case of $bg\to t W^-$.} and therefore also in the product of the two beam functions. In fact, the soft function for color-singlet final states becomes trivial to all orders in $\alpha_s$ with the analytic regulator, as the soft integrals are scaleless. Moreover, there is no double-counting between the soft function and the beam function in this regulator; otherwise, a soft-bin subtraction would be needed to generate the correct result in the effective theory~\cite{Chiu:2009yx}. Since the virtual and real integrals are from $-\infty$ to $+\infty$, the purely virtual integrals are scaleless and should vanish. Only integrals with at least one real emission survive. As the limits $\alpha_1\to 0$ and $\epsilon\to 0$ do not generally commute, in this context, we assume that the $\alpha_1\to 0$ limit is performed prior to the $\epsilon\to 0$ expansion. Specifically, we choose the order:
\begin{equation}
\alpha_1\to 0, \quad \epsilon\to0\,.\label{eq:order4regulators}
\end{equation}
In other words, we assume $|\alpha_1|\ll |\epsilon|$.

\section{One-loop soft function at higher orders in $\epsilon$\label{sec:softfunction}}

We are now in a position to consider the one-loop soft function at higher orders in the dimensional regulator $\epsilon$ expansion. In SCET, the virtual soft integrals are scaleless and thus vanish in dimensional regularization with the analytic rapidity regulator (eq.~\eqref{eq:analyticrapidityregulation}). Therefore, we only need to consider the single soft gluon real emission in the $2\to n+1$ process $r=(\ident_1,\ident_2,\ldots,\ident_{n+3})$, where we assume $\ident_{n+3}=g$, with momentum $k_g$ in the soft region. The reduced partonic subprocess, obtained by simply removing the soft parton $\ident_{n+3}=g$ from $r$, is denoted as
\begin{equation}
r^{\hspace{-1mm}{\scriptsize \backslashover{n+3}}}=\left(\ident_1, \ldots,\ident_{n+2},\remove{\ident}{0.2}_{n+3}\right)\,,
\end{equation}
and coincides with the original Born-level subprocess. In the soft limit of $\ident_{n+3}=g$, the real amplitude factorizes into the reduced Born amplitude multiplied by an eikonal factor. If the soft gluon with momentum $k_g$ is emitted from an external leg $\ident_j$ ($j\leq \nheavy$) with momentum $k_j$, the real emission amplitude takes the soft or eikonal approximation form:~\footnote{We generally only consider elementary particles here, as the IR structure might differ significantly for composite particles. A particular example is P-wave quarkonium production~\cite{AH:2024ueu}.} 
\begin{align}
\lim_{k_g \rightarrow 0}{\ampnpot(r)} = g_s\frac{k_j \cdot \varepsilon_{\lambda_{g}}^*(k_g)}{k_j\cdot k_g} \Qop(\ident_j)  \ampnt(r^{\hspace{-1mm}{\scriptsize \backslashover{n+3}}}),
\end{align}
where $g_s=\sqrt{4\pi\alpha_s}$ is the strong coupling, and $\Qop(\ident) $ represents the color generator associated with the particle $\ident$:
\begin{align}\label{SUN}
\Qop(\ident) = \{t^a\}_{a=1}^{8} ,\quad  \{-t^{aT}\}_{a=1}^8, \quad \{T^a\}_{a=1}^8 \quad \ident\in {\irrep{3},\irrepbar{3},\irrep{8}}  
\end{align}
with $t^a$ and $T^a$ being the SU(3) generators in the fundamental and adjoint representations, respectively. We take the final-state quarks or initial-state antiquarks as $\irrep{3}$, while initial-state quarks and final-state antiquarks belong to $\irrepbar{3}$. The adjoint representation matrix elements are given by $T^a_{bc}=-if_{abc}$. Color conservation imposes 
\begin{equation}
\sum_{i=1}^{\nheavy+2}{\Qop(\ident_i)}=\vec{0}\label{eq:colorconservation}
\end{equation}
in the (reduced) Born-level partonic process. In general, the soft limit of the squared real amplitude can be expressed as:
\begin{eqnarray}
\lim_{k_g\to 0}{\ampsqnpot(r)}&=&g_s^2\mathop{\sum_{k,l=1}}_{k\leq l}^{\nheavy+2}{\ampsqnt_{{\rm soft}, kl}(r)}\,,
\end{eqnarray}
where
\begin{eqnarray}
\ampsqnt_{{\rm soft}, kl}(r)&=&\frac{k_k\mydot k_l}{k_k\mydot k_g k_l\mydot k_g}\ampsqnt_{kl}(r^{\hspace{-1mm}{\scriptsize \backslashover{n+3}}})\,.\label{eq:softampsq}
\end{eqnarray}

The one-loop azimuthal-dependent (bare) TMD soft function is given by
\begin{equation}
\bm{S}_{\perp,\Ione\Itwo,l^\prime l}^{(1)}\left(\harpoon{b}_{T},\mu\right)=-\frac{\alpha_s}{2\pi} \sum_{k=1}^{\nheavy+2}{\sum_{l=k}^{\nheavy+2}{(2-\delta_{k,l})\bar{\mathcal{E}}_{\perp,kl}^{(m_k,m_l)}(\harpoon{b}_{T},\mu)\langle C_{l^\prime}|\Qop(\ident_k)\mydot\Qop(\ident_l)|C_l\rangle}}\,,\label{eq:TMD1Lsoftfuns}
\end{equation}
while its azimuthally-averaged counterpart is
\begin{equation}
\bm{S}_{\perp,\Ione\Itwo,l^\prime l}^{(1)}\left(b_{T},\mu\right)=-\frac{\alpha_s}{2\pi} \sum_{k=1}^{\nheavy+2}{\sum_{l=k}^{\nheavy+2}{(2-\delta_{k,l})\bar{\mathcal{E}}_{\perp,kl}^{(m_k,m_l)}(b_{T},\mu)\langle C_{l^\prime}|\Qop(\ident_k)\mydot\Qop(\ident_l)|C_l\rangle}}\,.\label{eq:TMD1Lsoftfunsazimuthaveraged}
\end{equation}
As is well known, the one-loop soft functions only exhibit a dipole color structure. The TMD eikonal integrals in $d=4-2\epsilon$ spacetime dimensions are given by
\begin{eqnarray}
&&\bar{\mathcal{E}}_{\perp,kl}^{(m_k,m_l)}(\harpoon{b}_{T},\mu)=8\pi^2\mu^{2\epsilon}\int{\frac{d^dk_g}{\left(2\pi\right)^d}2\pi \delta^{(+)}(k_g^2)\frac{k_k\cdot k_l}{k_k\cdot k_g k_l\cdot k_g}e^{-i\harpoon{k}_{g,T}\cdot \harpoon{b}_{T}}\left(\frac{\nu_1}{n\cdot k_g}\right)^{\alpha_1}},\nonumber\\
&&\bar{\mathcal{E}}_{\perp,kl}^{(m_k,m_l)}(b_{T},\mu)=\int{\frac{d\Omega_b^{(d-2)}}{\Omega_b^{(d-2)}}\bar{\mathcal{E}}_{\perp,kl}^{(m_k,m_l)}(\harpoon{b}_{T},\mu)}\,.
\label{eq:TMDeikonal}
\end{eqnarray}
In the definition of the TMD eikonal integral in eq.~\eqref{eq:TMDeikonal}, we have employed the (analytic) rapidity regulator~\cite{Becher:2010tm,Becher:2011dz} to regulate divergences arising when $k_g$ becomes collinear to either the first beam, $n$, or the second beam, $\bar{n}$. If an integral does not exhibit rapidity divergences, we can simply set the corresponding rapidity regulator $\alpha_1$ to zero from the outset. The rapidity regulator is necessary only when $k,l\in\left\{1,2\right\}$, as we consider here the case where all final-state colored particles at the Born level are massive.

In light-cone coordinates (cf. section \ref{sec:factorization}), it is convenient to integrate over 
\begin{equation}
\int{dk_{g,+}\delta^{(+)}(k_{g,+}k_{g,-}-k_{g,T}^2)}=\frac{1}{k_{g,-}}\,
\end{equation}
with $k_{g,+}=k_{g,T}^2/k_{g,-}$. The integration range for $k_{g,-}$ extends from $0$ to $+\infty$ due to the $(+)$ superscript in the Dirac delta function. Substituting this into equation \eqref{eq:TMDeikonal}, we obtain
\begin{equation}
\bar{\mathcal{E}}_{\perp,kl}^{(m_k,m_l)}(\harpoon{b}_{T},\mu)=8\pi^2\mu^{2\epsilon}\int{\frac{dk_{g,-}d^{d-2}\harpoon{k}_{g,T}}{\left(2\pi\right)^{d-1}}\frac{1}{2k_{g,-}}\frac{k_k\cdot k_l}{k_k\cdot k_g k_l\cdot k_g}e^{-i\harpoon{k}_{g,T}\cdot \harpoon{b}_{T}}\left(\frac{\nu_1}{k_{g,-}}\right)^{\alpha_1}}\,.\label{eq:TMDeikonal2}
\end{equation}
A similar expression holds if we swap $k_{g,-}$ and $k_{g,+}$. 

In the following, we discuss the computation of the TMD eikonal integrals in eq.~\eqref{eq:TMDeikonal} for four different cases:
\begin{enumerate}
    \item Two massless particles ($m_k,m_l=0$; cf. section \ref{sec:TMD1Leikonalmassless}),
    \item Self-massive case ($k=l,m_k=m_l\neq 0$; cf. section \ref{sec:TMD1Leikonalselfmassive}),
    \item One massless and one massive particles ($m_k=0,m_l\neq 0$; cf. section \ref{sec:TMD1Leikonalmasslessmassive}),
    \item Two distinct massive particles ($k\neq l, m_k,m_l\neq0$; cf. section \ref{sec:TMD1Leikonaltwomassive}).
\end{enumerate}
These cases are presented in roughly increasing order of complexity. In each case, we derive both the azimuthal-angle-averaged and azimuthal-angle-dependent results.

\subsection{Two massless case\label{sec:TMD1Leikonalmassless}}

The simplest case is the massless scenario with $m_k=m_l=0$. In this case, the eikonal integral can be nonzero only when $k=1$ and $l=2$; otherwise, if $k=l$, $k_k\cdot k_l=0$, leading to a vanishing integral. For $(k,l)=(1,2)$, the integral from eq.~\eqref{eq:TMDeikonal2} is
\begin{equation}
\begin{aligned}
\bar{\mathcal{E}}_{\perp,12}^{(0,0)}(\harpoon{b}_{T},\mu)=&8\pi^2 \mu^{2\epsilon}\int{\frac{dk_{g,-}d^{d-2}\harpoon{k}_{g,T}}{\left(2\pi\right)^{d-1}}\frac{1}{2k_{g,-}}\frac{k_1\cdot k_2}{k_1\cdot k_g k_2\cdot k_g}e^{-i\harpoon{k}_{g,T}\cdot \harpoon{b}_{T}}\left(\frac{\nu_1}{k_{g,-}}\right)^{\alpha_1}}\\
=&8\pi^2 \mu^{2\epsilon}\nu_1^{\alpha_1}\int{\frac{dk_{g,-}d^{d-2}\harpoon{k}_{g,T}}{\left(2\pi\right)^{d-1}}\frac{1}{k_{g,-}^{1+\alpha_1} k_{g,T}^2}e^{-i\harpoon{k}_{g,T}\cdot \harpoon{b}_{T}}}\\
=&8\pi^2 \mu^{2\epsilon}\nu_1^{\alpha_1}\int_0^{+\infty}{\frac{dk_{g,-}}{\left(2\pi\right)^{d-1}}\frac{1}{k_{g,-}^{1+\alpha_1}}\int{d^{d-2}\harpoon{k}_{g,T}\frac{1}{k_{g,T}^2}e^{-i\harpoon{k}_{g,T}\cdot \harpoon{b}_{T}}}}\\
=&2^{4-\epsilon}\pi^{2-\epsilon} b_T^\epsilon \mu^{2\epsilon}\nu_1^{\alpha_1}\int_0^{+\infty}{\frac{dk_{g,-}}{\left(2\pi\right)^{d-1}}\frac{1}{k_{g,-}^{1+\alpha_1}}\int_0^{+\infty}{dk_{g,T}k_{g,T}^{-1-\epsilon}J_{-\epsilon}\left(b_Tk_{g,T}\right)}}\\
=&\pi^{\epsilon-1} \Gamma\left(-\epsilon\right)b_T^{2\epsilon} \mu^{2\epsilon}\nu_1^{\alpha_1}\int_0^{+\infty}{dk_{g,-}\frac{1}{k_{g,-}^{1+\alpha_1}}}=0\,.
\end{aligned}
\end{equation}
The integral vanishes because it is scaleless, as expected. Consequently, the azimuthal-angle-averaged function $\bar{\mathcal{E}}_{\perp,12}^{(0,0)}(b_{T},\mu)$ is also zero. However, this result may differ in alternative rapidity regularization schemes, such as the ``delta" regulator~\cite{Chiu:2009yx,Echevarria:2011epo} or the ``exponential" regulator~\cite{Li:2016axz,Li:2016ctv}.

\subsection{Massive self-eikonal case\label{sec:TMD1Leikonalselfmassive}}

In this section, we consider the massive self-eikonal case with $k=l\geq 3$ and $m_k=m_l\neq 0$. Since there is no rapidity divergence in this case, we can safely set $\alpha_1=0$ at the integrand level. The eikonal integral can then be computed directly:
\begin{eqnarray}
&&\bar{\mathcal{E}}_{\perp,ll}^{(m_l,m_l)}(\harpoon{b}_{T},\mu)\nonumber\\
&=&8\pi^2m_l^2\mu^{2\epsilon}\int{\frac{dk_{g,-}d^{d-2}\harpoon{k}_{g,T}}{\left(2\pi\right)^{d-1}}\frac{1}{2k_{g,-}}\frac{1}{\left(k_l\cdot k_g\right)^2}e^{-i\harpoon{k}_{g,T}\cdot \harpoon{b}_{T}}}\nonumber\\
&=&8\pi^2m_l^2\mu^{2\epsilon}\int{\frac{dk_{g,-}d^{d-2}\harpoon{k}_{g,T}}{\left(2\pi\right)^{d-1}}\frac{2k_{g,-}}{\left(k_{l,-}k_{g,T}^2+k_{g,-}^2k_{l,+}-2k_{g,-}\harpoon{k}_{g,T}\cdot \harpoon{k}_{l,T}\right)^2}e^{-i\harpoon{k}_{g,T}\cdot \harpoon{b}_{T}}}\nonumber\\
&=&8\pi^2m_l^2\mu^{2\epsilon}\int{\frac{dk_{g,-}}{\left(2\pi\right)^{d-1}}\frac{2k_{g,-}}{k_{l,-}^2}e^{-i\frac{k_{g,-}}{k_{l,-}}\harpoon{k}_{l,T}\cdot \harpoon{b}_{T}}\int{d^{d-2}\harpoon{K}_{T}\frac{1}{\left(K_{T}^2+m_l^2\frac{k_{g,-}^2}{k_{l,-}^2}\right)^2}e^{-i\harpoon{K}_{T}\cdot \harpoon{b}_{T}}}}\nonumber\\
&=&2m_l^2\left(2\pi\right)^{\epsilon}b_T^\epsilon \mu^{2\epsilon}\int_0^{+\infty}{dk_{g,-}\frac{2k_{g,-}}{k_{l,-}^2}e^{-i\frac{k_{g,-}}{k_{l,-}}\harpoon{k}_{l,T}\cdot \harpoon{b}_{T}}\int_0^{+\infty}{dK_{T}\frac{K_T^{1-\epsilon} J_{-\epsilon}\left(b_TK_T\right)}{\left(K_{T}^2+m_l^2\frac{k_{g,-}^2}{k_{l,-}^2}\right)^2}}}\nonumber\\
&=&2\left(\frac{m_l}{k_{l,-}}\right)^{1-\epsilon}\left(2\pi\right)^{\epsilon}b_T^{1+\epsilon} \mu^{2\epsilon}\int_0^{+\infty}{dk_{g,-}k_{g,-}^{-\epsilon}K_{-1-\epsilon}\left(b_Tm_l\frac{k_{g,-}}{k_{l,-}}\right)e^{-i\frac{k_{g,-}}{k_{l,-}}\harpoon{k}_{l,T}\cdot \harpoon{b}_{T}}}\nonumber\\
&=&\pi^\ep\left(\mu b_T\right)^{2\epsilon}\left[\Gamma\left(-\epsilon\right){}_2\hspace{-0.7mm}F_1\left(1,-\epsilon;\frac{1}{2};-\frac{\left(\harpoon{k}_{l,T}\cdot \harpoon{b}_{T}\right)^2}{b_T^2m_l^2}\right)\right.\nonumber\\
&&\left.-i\sqrt{\pi}\left(b_Tm_l\right)^{-2\epsilon}\left(\harpoon{k}_{l,T}\cdot \harpoon{b}_{T}\right)\left(\left(\harpoon{k}_{l,T}\cdot \harpoon{b}_{T}\right)^2+\left(b_Tm_l\right)^2\right)^{-\frac{1}{2}+\epsilon}\Gamma\left(\frac{1}{2}-\epsilon\right)\right]\nonumber\\
&=&\frac{(4\pi)^\epsilon}{\Gamma(1-\epsilon)}\Bigg\{-\frac{1}{\epsilon}-L_\perp+y\log{\left(\frac{1-y}{1+y}\right)}-\epsilon\bigg[\frac{L_\perp^2}{2!}+\zeta_2-L_\perp y\log{\left(\frac{1-y}{1+y}\right)}\nonumber\\
&&+\frac{y}{2}\left(\log^2{\left(\frac{1-y}{2}\right)}-\log^2{\left(\frac{1+y}{2}\right)}\right)+y\left(\mathrm{Li}_2\left(\frac{1-y}{2}\right)-\mathrm{Li}_2\left(\frac{1+y}{2}\right)\right)\bigg]\nonumber\\
&&-\epsilon^2\bigg[\frac{L_\perp^3}{3!}+L_\perp \zeta_2+\frac{2}{3}\zeta_3-\frac{L_\perp^2}{2!}y\log{\left(\frac{1-y}{1+y}\right)}+\frac{1}{2}L_\perp y\left(\log^2{\left(\frac{1-y}{2}\right)}-\log^2{\left(\frac{1+y}{2}\right)}\right)\nonumber\\
&&-\frac{1}{6}y\log{\left(\frac{1-y}{1+y}\right)}\left(\log^2{\left(\frac{1-y}{2}\right)}+\log^2{\left(\frac{1+y}{2}\right)}+4\log{\left(\frac{1-y}{2}\right)}\log{\left(\frac{1+y}{2}\right)}\right)\nonumber\\
&&+L_\perp y \left(\mathrm{Li}_2\left(\frac{1-y}{2}\right)-\mathrm{Li}_2\left(\frac{1+y}{2}\right)\right)+2y\left(\mathrm{Li}_3\left(\frac{1+y}{2}\right)-\mathrm{Li}_3\left(\frac{1-y}{2}\right)\right)\bigg]\nonumber\\
&&-\epsilon^3\bigg[\frac{L_\perp^4}{4!}+\frac{1}{2}L_\perp^2\zeta_2+\frac{2}{3}L_\perp \zeta_3+\frac{7}{4}\zeta_4-y\left(\frac{L_\perp^3}{3!}+\frac{5}{3}\zeta_3\right)\log{\left(\frac{1-y}{1+y}\right)}\nonumber\\
&&+\frac{1}{4}L_\perp^2y\left(\log^2{\left(\frac{1-y}{2}\right)}-\log^2{\left(\frac{1+y}{2}\right)}\right)+\frac{1}{24}\left(\log{\left(\frac{1-y^2}{4}\right)}-4L_\perp\right) \nonumber\\
&&\times y \log{\left(\frac{1-y}{1+y}\right)}\left(\log^2{\left(\frac{1-y}{2}\right)}+\log^2{\left(\frac{1+y}{2}\right)}+4\log{\left(\frac{1-y}{2}\right)}\log{\left(\frac{1+y}{2}\right)}\right)\nonumber\\
&&+y\left(\frac{L_\perp^2}{2!}+\zeta_2\right)\left(\mathrm{Li}_2\left(\frac{1-y}{2}\right)-\mathrm{Li}_2\left(\frac{1+y}{2}\right)\right)\nonumber\\
&&+y\left(\left(2L_\perp-\log{\left(\frac{1-y}{2}\right)}\right)\mathrm{Li}_3\left(\frac{1+y}{2}\right)-\left(2L_\perp-\log{\left(\frac{1+y}{2}\right)}\right)\mathrm{Li}_3\left(\frac{1-y}{2}\right)\right)\nonumber\\
&&+y\left(\mathrm{S}_{2,2}\left(\frac{1-y}{2}\right)-\mathrm{S}_{2,2}\left(\frac{1+y}{2}\right)\right)+2y\left(\mathrm{Li}_4\left(\frac{1-y}{2}\right)-\mathrm{Li}_4\left(\frac{1+y}{2}\right)\right)\bigg]\nonumber\\
&&-i\pi \frac{\harpoon{k}_{l,T}\cdot \harpoon{b}_{T}}{|\harpoon{k}_{l,T}\cdot \harpoon{b}_{T}|}y\bigg[1+\epsilon\left(L_\perp-\log{\left(\frac{1-y^2}{4}\right)}\right)\nonumber\\
&&+\epsilon^2\left(\frac{L_\perp^2}{2!}+2\zeta_2-L_\perp\log{\left(\frac{1-y^2}{4}\right)}+\frac{1}{2}\log^2{\left(\frac{1-y^2}{4}\right)}\right)\nonumber\\
&&+\epsilon^3\bigg(\frac{L_\perp^3}{3!}+2L_\perp \zeta_2+\frac{8}{3}\zeta_3-\left(\frac{L_\perp^2}{2!}+2\zeta_2\right)\log{\left(\frac{1-y^2}{4}\right)}\nonumber\\
&&+\frac{1}{2}L_\perp\log^2{\left(\frac{1-y^2}{4}\right)}-\frac{1}{6}\log^3{\left(\frac{1-y^2}{4}\right)}\bigg)\bigg]\Bigg\}+\mathcal{O}(\epsilon^4)\,,\label{eq:selfeikonal}
\end{eqnarray}
where, in the third equation, we have introduced the new variable~\cite{Catani:2023tby}
\begin{eqnarray}
\harpoon{K}_{T}&\equiv&\harpoon{k}_{g,T}-\frac{k_{g,-}}{k_{l,-}}\harpoon{k}_{l,T}\,,\label{eq:KTdef}
\end{eqnarray}
and, in the fourth equation, we have used eq.~\eqref{eq:solidangleFT}. The function $K_\alpha()$ denotes the modified Bessel function of the second kind, $\zeta_n$ is the Riemann zeta function $\zeta(n)$, $\mathrm{Li}_n()$ is the classical polylogarithm, and $\mathrm{S}_{n,p}()$ represents Nielsen's generalized polylogarithm. Additionally, we introduce the new variable in eq.~\eqref{eq:selfeikonal}:
\begin{equation}
y\equiv \frac{|\harpoon{k}_{l,T}\cdot \harpoon{b}_{T}|}{\sqrt{\left(\harpoon{k}_{l,T}\cdot \harpoon{b}_{T}\right)^2+b_T^2m_l^2}}\,,\label{eq:ydefinition}
\end{equation}
which takes values in the range of $[0,1)$. An important observation in eq.~\eqref{eq:selfeikonal} is that the real part of the eikonal integral is symmetric under the replacement $y\to -y$, while the imaginary part remains invariant under the simultaneous transformation $y\to -y$ and 
\begin{equation}
\frac{\harpoon{k}_{l,T}\cdot \harpoon{b}_{T}}{|\harpoon{k}_{l,T}\cdot \harpoon{b}_{T}|}\to -\frac{\harpoon{k}_{l,T}\cdot \harpoon{b}_{T}}{|\harpoon{k}_{l,T}\cdot \harpoon{b}_{T}|}\,.
\end{equation}
The Laurent series expansion in $\epsilon$ of the hypergeometric function appearing in eq.~\eqref{eq:selfeikonal} was obtained using the {\sc\small Mathematica} package {\tt HypExp}~\cite{Huber:2005yg,Huber:2007dx}. In the last identity of eq.~\eqref{eq:selfeikonal}, we retain only terms up to $\mathcal{O}(\epsilon^3)$. Since the hypergeometric function and the Euler gamma function $\Gamma()$ are known for all orders in $\epsilon$, higher-order terms in $\epsilon$ can be obtained systematically without difficulty. An overall global factor $(4\pi)^\epsilon/\Gamma(1-\epsilon)$ has been factored out in the final expression, following the Binoth Les
Houches Accord~\cite{Binoth:2010xt,Alioli:2013nda}. This is because one-loop hard functions will be obtained from automated one-loop providers, which commonly employ the same global factor. Finally, if we assign a transcendentality weight of $-1$ to the dimensional regulator $\epsilon$, the $\epsilon$-expanded expression in the curly brackets on the right-hand side of eq.~\eqref{eq:selfeikonal} exhibits the uniform transcendental (UT) property, with a weight of unity.

The azimuthal-angle-averaged integral can be easily derived from the fifth equation in eq.~\eqref{eq:selfeikonal}:
\begin{eqnarray}
&&\bar{\mathcal{E}}_{\perp,ll}^{(m_l,m_l)}(b_{T},\mu)\nonumber\\
&=&\frac{2}{\Omega_b^{(d-2)}}\left(\frac{m_l}{k_{l,-}}\right)^{1-\epsilon}\left(2\pi\right)^{\epsilon}b_T^{1+\epsilon} \mu^{2\epsilon}\int_0^{+\infty}{dk_{g,-}k_{g,-}^{-\epsilon}K_{-1-\epsilon}\left(b_Tm_l\frac{k_{g,-}}{k_{l,-}}\right)\int{d\Omega^{(d-2)}_b e^{-i\frac{k_{g,-}}{k_{l,-}}\harpoon{k}_{l,T}\cdot \harpoon{b}_{T}}}}\nonumber\\
&=&\frac{m_l}{k_{l,-}}\left(\frac{k_{l,T}}{m_l}\right)^\epsilon 2\pi^\epsilon\Gamma\left(1-\epsilon\right)b_T^{1+2\epsilon} \mu^{2\epsilon}\int_0^{+\infty}{dk_{g,-}K_{-1-\epsilon}\left(b_Tm_l\frac{k_{g,-}}{k_{l,-}}\right)J_{-\epsilon}\left(b_Tk_{l,T}\frac{k_{g,-}}{k_{l,-}}\right)}\nonumber\\
&=&\pi^\epsilon\left(b_T\mu\right)^{2\epsilon}\Gamma\left(-\epsilon\right){}_2\hspace{-0.7mm}F_1\left(1,-\epsilon;1-\epsilon;-\frac{k_{l,T}^2}{m_l^2}\right)\nonumber\\
&=&\pi^\epsilon\left(b_T\mu\right)^{2\epsilon}\Gamma\left(-\epsilon\right)\left[1-\sum_{i=1}^{+\infty}{\epsilon^i \mathrm{Li}_i\left(-\frac{k_{l,T}^2}{m_l^2}\right)}\right]\nonumber\\
&=&\frac{(4\pi)^\epsilon}{\Gamma(1-\epsilon)}\Bigg[-\frac{1}{\epsilon}-L_\perp-\log{\left(\frac{k_{l,+}k_{l,-}}{m_l^2}\right)}-\epsilon\bigg(\frac{L_\perp^2}{2!}+\zeta_2+L_\perp \log{\left(\frac{k_{l,+}k_{l,-}}{m_l^2}\right)}-\mathrm{Li}_2\left(-\frac{k_{l,T}^2}{m_l^2}\right)\bigg)\nonumber\\
&&-\epsilon^2\bigg(\frac{L_\perp^3}{3!}+L_\perp \zeta_2+\frac{2}{3}\zeta_3+\left(\frac{L_\perp^2}{2!}+\zeta_2\right)\log{\left(\frac{k_{l,+}k_{l,-}}{m_l^2}\right)}-L_\perp \mathrm{Li}_2\left(-\frac{k_{l,T}^2}{m_l^2}\right)-\mathrm{Li}_3\left(-\frac{k_{l,T}^2}{m_l^2}\right)\bigg)\nonumber\\
&&-\epsilon^3\bigg(\frac{L_\perp^4}{4!}+\frac{1}{2}L_\perp^2\zeta_2+\frac{2}{3}L_\perp\zeta_3+\frac{7}{4}\zeta_4+\left(\frac{L_\perp^3}{3!}+L_\perp \zeta_2+\frac{2}{3}\zeta_3\right)\log{\left(\frac{k_{l,+}k_{l,-}}{m_l^2}\right)}\nonumber\\
&&-\left(\frac{L_\perp^2}{2!}+\zeta_2\right)\mathrm{Li}_2\left(-\frac{k_{l,T}^2}{m_l^2}\right)-L_\perp\mathrm{Li}_3\left(-\frac{k_{l,T}^2}{m_l^2}\right)-\mathrm{Li}_4\left(-\frac{k_{l,T}^2}{m_l^2}\right)\bigg)\Bigg]+\mathcal{O}(\epsilon^4)\,,\label{eq:selfeikonalavg}
\end{eqnarray}
where $\mathrm{Li}_1(x)=-\log{\left(1-x\right)}$. In contrast to the azimuthal-angle-dependent case in eq.~\eqref{eq:selfeikonal}, the azimuthally averaged eikonal integral has a vanishing imaginary part. Equation \eqref{eq:selfeikonalavg} agrees with eq.(95) in ref.~\cite{Catani:2023tby}, up to $\mathcal{O}(\epsilon^1)$, apart from a global factor of $8\pi^2\mu^{2\epsilon}/\left(2\pi\right)^{d-1}$. Alternatively, one can use the sixth equation in eq.~\eqref{eq:selfeikonal} and the series representation of the hypergeometric function
\begin{eqnarray}
{}_2\hspace{-0.7mm}F_1(a,b;c;z)&=&\sum_{i=0}^{+\infty}{\frac{(a)_i(b)_i}{(c)_i}\frac{z^i}{i!}}\,,\label{eq:series42F1}
\end{eqnarray}
where $(a)_i\equiv\Gamma(a+i)/\Gamma(a)$ is the Pochhammer symbol,
to directly perform the integration over $d\Omega_b^{(d-2)}$. The second term vanishes after integrating over the azimuthal angle $d\Omega_b^{(d-2)}$ because the imaginary part is proportional to the sign of $\harpoon{k}_{l,T}\cdot \harpoon{b}_T$. As expected, eq.~\eqref{eq:selfeikonalavg} also satisfies the UT property.

\subsection{One-massless-one-massive case\label{sec:TMD1Leikonalmasslessmassive}}

In this section, we consider the one-massless, one-massive case. This scenario can only occur when the massless external particle is in the initial state and the massive external particle is in the final state. In other words, we have $k\leq 2$ ($m_k=0$) and $3\leq l\leq \nheavy+2$ ($m_l\neq 0$). 

Let us first consider the case where $k=1$ and $l\geq 3$, ensuring that $m_k=0$ and $m_l\neq 0$. The full azimuthal-angle-dependent TMD eikonal integral is given by
\begin{eqnarray}
&&\bar{\mathcal{E}}_{\perp,1l}^{(0,m_l)}(\harpoon{b}_{T},\mu)\nonumber\\
&=&8\pi^2\mu^{2\epsilon}\int{\frac{d^dk_g}{\left(2\pi\right)^d}2\pi \delta^{(+)}(k_g^2)\frac{k_1\cdot k_l}{k_1\cdot k_g k_l\cdot k_g}e^{-i\harpoon{k}_{g,T}\cdot \harpoon{b}_{T}}\left(\frac{\nu_1}{n\cdot k_g}\right)^{\alpha_1}}\nonumber\\
&=&4\pi^2\mu^{2\epsilon}\nu_1^{\alpha_1}k_{l,-}\int{\frac{dk_{g,-}d^{d-2}\harpoon{k}_{g,T}}{\left(2\pi\right)^{d-1}}\frac{1}{k_{g,-}^{\alpha_1+2}}\frac{1}{k_l\cdot k_g}e^{-i\harpoon{k}_{g,T}\cdot \harpoon{b}_{T}}}\nonumber\\
&=&8\pi^2\mu^{2\epsilon}\nu_1^{\alpha_1}k_{l,-}\int{\frac{dk_{g,-}d^{d-2}\harpoon{k}_{g,T}}{\left(2\pi\right)^{d-1}}\frac{1}{k_{g,-}^{\alpha_1+1}}\frac{1}{k_{l,-}k_{g,T}^2+k_{l,+}k_{g,-}^2-2k_{g,-}\harpoon{k}_{l,T}\cdot \harpoon{k}_{g,T}}e^{-i\harpoon{k}_{g,T}\cdot \harpoon{b}_{T}}}\nonumber\\
&=&8\pi^2\mu^{2\epsilon}\nu_1^{\alpha_1}\int_0^{+\infty}{\frac{dk_{g,-}}{\left(2\pi\right)^{d-1}}\frac{1}{k_{g,-}^{\alpha_1+1}}e^{-i\frac{k_{g,-}}{k_{l,-}}\harpoon{k}_{l,T}\cdot \harpoon{b}_T}\int{d^{d-2}\harpoon{K}_{T}\frac{1}{K_{T}^2+m_l^2\frac{k_{g,-}^2}{k_{l,-}^2}}e^{-i\harpoon{K}_{T}\cdot \harpoon{b}_{T}}}}\nonumber\\
&=&2(2\pi)^\epsilon b_T^\epsilon \mu^{2\epsilon}\nu_1^{\alpha_1}\int_0^{+\infty}{dk_{g,-}\frac{1}{k_{g,-}^{\alpha_1+1}}e^{-i\frac{k_{g,-}}{k_{l,-}}\harpoon{k}_{l,T}\cdot \harpoon{b}_T}\int_0^{+\infty}{dK_T}\frac{K_T^{1-\epsilon}J_{-\epsilon}(b_TK_T)}{K_{T}^2+m_l^2\frac{k_{g,-}^2}{k_{l,-}^2}}}\nonumber\\
&=&2(2\pi)^\epsilon b_T^\epsilon \mu^{2\epsilon}\nu_1^{\alpha_1}\left(\frac{k_{l,-}}{m_l}\right)^\epsilon\int_0^{+\infty}{dk_{g,-}\frac{1}{k_{g,-}^{\alpha_1+1+\epsilon}}K_{\epsilon}\left(b_Tm_l\frac{k_{g,-}}{k_{l,-}}\right)e^{-i\frac{k_{g,-}}{k_{l,-}}\harpoon{k}_{l,T}\cdot \harpoon{b}_T}}\nonumber\\
&=&\pi^\epsilon \left(\mu b_T\right)^{2\epsilon}\left(\frac{b_T m_l \nu_1}{2k_{l,-}}\right)^{\alpha_1}\Bigg\{\frac{1}{2}\Gamma\left(-\frac{\alpha_1}{2}\right)\Gamma\left(-\frac{\alpha_1}{2}-\epsilon\right)\nonumber\\
&&\times {}_2\hspace{-0.7mm}F_1\left(-\frac{\alpha_1}{2},-\frac{\alpha_1}{2}-\epsilon;\frac{1}{2};-\frac{\left(\harpoon{k}_{l,T}\cdot \harpoon{b}_T\right)^2}{b_T^2 m_l^2}\right)-i\frac{\Gamma\left(\frac{1}{2}-\frac{\alpha_1}{2}\right)\Gamma\left(\frac{1}{2}-\frac{\alpha_1}{2}-\epsilon\right)}{\alpha_1\left(\alpha_1+2\epsilon\right)b_Tm_l \left(\harpoon{k}_{l,T}\cdot \harpoon{b}_T\right)}\nonumber\\
&&\times\Bigg[\left(\left(\harpoon{k}_{l,T}\cdot \harpoon{b}_T\right)^2+b_T^2m_l^2\right){}_2\hspace{-0.7mm}F_1\left(\frac{1-\alpha_1}{2},\frac{1-\alpha_1-2\epsilon}{2};-\frac{1}{2};-\frac{\left(\harpoon{k}_{l,T}\cdot \harpoon{b}_T\right)^2}{b_T^2 m_l^2}\right)\nonumber\\
&&-\left(2\left(1-\alpha_1-\epsilon\right)\left(\harpoon{k}_{l,T}\cdot \harpoon{b}_T\right)^2+b_T^2m_l^2\right){}_2\hspace{-0.7mm}F_1\left(\frac{1-\alpha_1}{2},\frac{1-\alpha_1-2\epsilon}{2};\frac{1}{2};-\frac{\left(\harpoon{k}_{l,T}\cdot \harpoon{b}_T\right)^2}{b_T^2 m_l^2}\right)\Bigg]\Bigg\}\,,\nonumber\\\label{eq:E1l0ml}
\end{eqnarray}
where, in the fourth equality, we have used the definition of $\harpoon{K}_T$ given in eq.~\eqref{eq:KTdef}.

The integral present in eq.~\eqref{eq:E1l0ml} exhibits divergences both in $\alpha_1$ and $\epsilon$. However, since the final soft function remains finite in the limit $\alpha_1\to 0$, the apparent rapidity divergences in $\bar{\mathcal{E}}_{\perp,kl}^{(0,m_l)}(\harpoon{b}_{T},\mu)$ for $k=1$ or $2$ can be subtracted by introducing an $l$ independent integral $\bar{\mathcal{E}}_{\perp,k}(\harpoon{b}_T,\mu)$, defined as
\begin{equation}
\bar{\mathcal{E}}_{\perp,k}(\harpoon{b}_{T},\mu)\equiv 8\pi^2\mu^{2\epsilon}\int{\frac{d^dk_g}{(2\pi)^d}2\pi \delta^{(+)}(k_g^2)\frac{k_1\cdot k_2}{k_k\cdot k_g (k_1+k_2)\cdot k_g}e^{-i\harpoon{k}_{g,T}\cdot \harpoon{b}_T}\left(\frac{\nu_1}{n\cdot k_g}\right)^{\alpha_1}}\,.\label{eq:TMDeikonalspecial}
\end{equation}
From this definition, it follows that
\begin{equation}
\bar{\mathcal{E}}_{\perp,1}(\harpoon{b}_{T},\mu)+\bar{\mathcal{E}}_{\perp,2}(\harpoon{b}_{T},\mu)=\bar{\mathcal{E}}_{\perp,12}^{(0,0)}(\harpoon{b}_{T},\mu)=0\,,\label{eq:E1bTpE2bT}
\end{equation}
where $\bar{\mathcal{E}}_{\perp,12}^{(0,0)}(\harpoon{b}_{T},\mu)$ is zero in our rapidity and dimensional regularization scheme because it is a scaleless integral (cf. section \ref{sec:TMD1Leikonalmassless}). Now, if we apply the following replacements:
\begin{equation}
\bar{\mathcal{E}}_{\perp,kl}^{(0,m_l)}(\harpoon{b}_{T},\mu)\to \bar{\mathcal{E}}_{\perp,kl}^{(0,m_l)}(\harpoon{b}_{T},\mu)-\bar{\mathcal{E}}_{\perp,k}(\harpoon{b}_{T},\mu),\quad {\rm for}~1\leq k\leq 2, \; 3\leq l\leq \nheavy+2\,,\label{eq:Etrickrep}
\end{equation}
the soft function $\bm{S}_{\perp,\Ione\Itwo,l^\prime l}^{(1)}\left(\harpoon{b}_{T},\mu\right)$ (cf. eq.~\eqref{eq:TMD1Lsoftfuns}) remains unchanged due to color conservation eq.~\eqref{eq:colorconservation}.
This can be easily understood by noting that
\begin{equation}
\begin{aligned}
-\sum_{k=1}^{2}{\sum_{l=3}^{\nheavy+2}{\bar{\mathcal{E}}_{\perp,k}(\harpoon{b}_{T},\mu)\Qop(\ident_k)\cdot \Qop(\ident_l)}}&=\sum_{k=1}^{2}{\bar{\mathcal{E}}_{\perp,k}(\harpoon{b}_{T},\mu)\Qop(\ident_k)\cdot \left(\Qop(\Ione)+\Qop(\Itwo)\right)}\\
&=\left(\bar{\mathcal{E}}_{\perp,1}(\harpoon{b}_{T},\mu)+\bar{\mathcal{E}}_{\perp,2}(\harpoon{b}_{T},\mu)\right)\left(C(\Ione)+\Qop(\Ione)\cdot \Qop(\Itwo)\right)\\
&=\bar{\mathcal{E}}_{\perp,12}^{(0,0)}(\harpoon{b}_{T},\mu)\left(C(\Ione)+\Qop(\Ione)\cdot \Qop(\Itwo)\right)=0\,.\label{eq:prooftrick}
\end{aligned}
\end{equation}
Here, $C(\ident)=\Qop(\ident)\cdot \Qop(\ident)$ is the QCD Casimir factor associated with the particle identity $\ident$. These factors are
\begin{equation}
\begin{aligned}
C(\ident)&=\left\{\begin{array}{ll}
C_F=(N_c^2-1)/(2N_c)=4/3, &~~{\rm if}~~\ident\in \irrep{3},\irrepbar{3}\\
C_A=N_c=3, &~~{\rm if}~~\ident\in \irrep{8}\\
\end{array}\right..
\end{aligned}
\end{equation}
We have used the fact that $C(\Ione)=C(\Itwo)$ in deriving the second identity in eq.~\eqref{eq:prooftrick}. The advantage of applying eq.~\eqref{eq:Etrickrep} is that it ensures the right-hand side of the equation is free of any rapidity divergence. Similar replacements and $l$-independent integral definitions can be made for the azimuthal-angle-averaged integrals, though we will not repeat the details here. This trick can also be generalized to higher orders in $\alpha_s$.

Due to the introduction of the asymmetric rapidity regulator between the two beams, the eikonal integrals with $k=2$ do not have a simple relation to those with $k=1$. However, after subtracting the rapidity divergences using eq.~\eqref{eq:Etrickrep}, the $k=2$ eikonal integrals can be directly obtained from the $k=1$ ones:
\begin{eqnarray}
\bar{\mathcal{E}}_{\perp,2l}^{(0,m_l)}(b_{T},\mu)-\bar{\mathcal{E}}_{\perp,2}(b_{T},\mu)&=&\left[\bar{\mathcal{E}}_{\perp,1l}^{(0,m_l)}(b_{T},\mu)-\bar{\mathcal{E}}_{\perp,1}(b_{T},\mu)\right]_{k_{l,-}\to k_{l,+}}\,,\label{eq:E2bTmE1bTazimuthavg}\\
\bar{\mathcal{E}}_{\perp,2l}^{(0,m_l)}(\harpoon{b}_{T},\mu)-\bar{\mathcal{E}}_{\perp,2}(\harpoon{b}_{T},\mu)&=&\left[\bar{\mathcal{E}}_{\perp,1l}^{(0,m_l)}(\harpoon{b}_{T},\mu)-\bar{\mathcal{E}}_{\perp,1}(\harpoon{b}_{T},\mu)\right]_{k_{l,-}\to k_{l,+}}\,.\label{eq:E2bTmE1bTfullazimuth}
\end{eqnarray}
Thus, we only need to evaluate the integrals for $k=1$ and $3\leq l\leq \nheavy+2$.

\subsubsection{Azimuthal-angle-averaged result}

Based on the right-hand side of the 6th identity in eq.~\eqref{eq:E1l0ml}, we can obtain the azimuthally-averaged eikonal integral as follows:
\begin{eqnarray}
&&\bar{\mathcal{E}}_{\perp,1l}^{(0,m_l)}(b_{T},\mu)\nonumber\\
&=&\frac{2(2\pi)^\epsilon b_T^\epsilon \mu^{2\epsilon}\nu_1^{\alpha_1}}{\Omega_b^{(d-2)}}\left(\frac{k_{l,-}}{m_l}\right)^\epsilon\int_0^{+\infty}{dk_{g,-}\frac{1}{k_{g,-}^{\alpha_1+1+\epsilon}}K_{\epsilon}\left(b_Tm_l\frac{k_{g,-}}{k_{l,-}}\right)\int{d\Omega_b^{(d-2)} e^{-i\frac{k_{g,-}}{k_{l,-}}\harpoon{k}_{l,T}\cdot \harpoon{b}_T}}}\nonumber\\
&=&2\pi^\epsilon\Gamma\left(1-\epsilon\right) \left(b_T \mu\right)^{2\epsilon}\nu_1^{\alpha_1}\left(\frac{k_{l,T}}{m_l}\right)^\epsilon\int_0^{+\infty}{dk_{g,-}\frac{1}{k_{g,-}^{\alpha_1+1}}K_{\epsilon}\left(b_Tm_l\frac{k_{g,-}}{k_{l,-}}\right)J_{-\epsilon}\left(b_Tk_{l,T}\frac{k_{g,-}}{k_{l,-}}\right)}\nonumber\\
&=&\frac{\pi^\epsilon}{2}\left(b_T\mu\right)^{2\epsilon}\left(\frac{b_Tm_l \nu_1}{2k_{l,-}}\right)^{\alpha_1}\Gamma\left(-\frac{\alpha_1}{2}\right)\Gamma\left(-\frac{\alpha_1}{2}-\epsilon\right){}_2\hspace{-0.7mm}F_1\left(-\frac{\alpha_1}{2},-\frac{\alpha_1}{2}-\epsilon; 1-\epsilon; -\frac{k_{l,T}^2}{m_l^2}\right)\,.\nonumber\\
\label{eq:E1leikonal0ml1}
\end{eqnarray}
The full expression, including all orders in the two regulators $\alpha_1$ and $\epsilon$, has been written in terms of hypergeometric functions and gamma functions. This should be expanded
according to the order defined in eq.~\eqref{eq:order4regulators} to obtain the Laurent series expansion of eq.~\eqref{eq:E1leikonal0ml1}. In the limit $\alpha_1\to 0$, we can use the series representation of the hypergeometric function given in eq.~\eqref{eq:series42F1}. It is sufficient to keep expansion terms in the rapidity regulator $\alpha_1$ up to $\mathcal{O}(\alpha_1)$, even for higher-order computations, because the soft function remains free of rapidity divergences with the analytic regulator. The hypergometric function in eq.~\eqref{eq:E1leikonal0ml1} can be expanded as:
\begin{eqnarray}
{}_2\hspace{-0.7mm}F_1\left(-\frac{\alpha_1}{2},-\frac{\alpha_1}{2}-\epsilon; 1-\epsilon; x\right)&=&1-\frac{\alpha_1}{2}\sum_{i=1}^{+\infty}{\frac{1}{i}\frac{(1)_i(-\epsilon)_i}{(1-\epsilon)_i}\frac{x^i}{i!}}+\mathcal{O}(\alpha_1^2)\nonumber\\
&=&1+\frac{\alpha_1}{2}\left[\frac{x}{1-\epsilon}{}_2\hspace{-0.7mm}F_1\left(1,1-\epsilon; 2-\epsilon; x\right)+\log{\left(1-x\right)}\right]+\mathcal{O}(\alpha_1^2)\nonumber\\
&=&1+\frac{\alpha_1}{2}\sum_{i=1}^{+\infty}{\epsilon^i\mathrm{Li}_{i+1}\left(x\right)}+\mathcal{O}(\alpha_1^2)\,,\label{eq:hypergeometric1F1exp1}
\end{eqnarray}
which gives us
\begin{align}
&\bar{\mathcal{E}}_{\perp,1l}^{(0,m_l)}(b_{T},\mu)\nonumber\\
=&\frac{(4\pi)^\epsilon}{\Gamma(1-\epsilon)}\Bigg\{\frac{1}{\alpha_1}\bigg[\frac{1}{\epsilon}+L_\perp+\epsilon\left(\frac{L_\perp^2}{2!}+\zeta_2\right)+\epsilon^2\left(\frac{L_\perp^3}{3!}+L_\perp \zeta_2+\frac{2}{3}\zeta_3\right)\nonumber\\
&+\epsilon^3\left(\frac{L_\perp^4}{4!}+\frac{1}{2}L_\perp^2\zeta_2+\frac{2}{3}L_\perp \zeta_3+\frac{7}{4}\zeta_4\right)\nonumber\\
&+\epsilon^4\left(\frac{L_\perp^5}{5!}+\frac{1}{6}L_\perp^3\zeta_2+\frac{1}{3}L_\perp^2\zeta_3+\frac{7}{4}L_\perp\zeta_4+\frac{2}{3}\zeta_2\zeta_3+\frac{2}{5}\zeta_5\right)+\mathcal{O}(\epsilon^5)\bigg]\nonumber\\
&-\frac{1}{2\epsilon^2}+\frac{\log{\left(\frac{\nu_1m_l}{\mu k_{l,-}}\right)}}{\epsilon}+\log{\left(\frac{\nu_1m_l}{\mu k_{l,-}}\right)}L_\perp+\frac{L_\perp^2}{4}+\frac{1}{2}\mathrm{Li}_2\left(-\frac{k_{l,T}^2}{m_l^2}\right)\nonumber\\
&+\epsilon\bigg[\log{\left(\frac{\nu_1m_l}{\mu k_{l,-}}\right)}\left(\frac{L_\perp^2}{2!}+\zeta_2\right)+\frac{L_\perp^3}{6}+\frac{1}{2}L_\perp\zeta_2+\frac{\zeta_3}{6}+\frac{1}{2}L_\perp \mathrm{Li}_2\left(-\frac{k_{l,T}^2}{m_l^2}\right)+\frac{1}{2}\mathrm{Li}_3\left(-\frac{k_{l,T}^2}{m_l^2}\right)\bigg]\nonumber\\
&+\epsilon^2\bigg[\log{\left(\frac{\nu_1m_l}{\mu k_{l,-}}\right)}\left(\frac{L_\perp^3}{3!}+L_\perp \zeta_2+\frac{2}{3}\zeta_3\right)+\frac{L_\perp^4}{16}+\frac{1}{2}L_\perp^2 \zeta_2+\frac{1}{2}L_\perp \zeta_3+\frac{7}{8}\zeta_4\nonumber\\
&+\left(\frac{L_\perp^2}{4}+\frac{\zeta_2}{2}\right)\mathrm{Li}_2\left(-\frac{k_{l,T}^2}{m_l^2}\right)+\frac{1}{2}L_\perp \mathrm{Li}_3\left(-\frac{k_{l,T}^2}{m_l^2}\right)+\frac{1}{2}\mathrm{Li}_4\left(-\frac{k_{l,T}^2}{m_l^2}\right)\bigg]\nonumber\\
&+\epsilon^3\bigg[\log{\left(\frac{\nu_1m_l}{\mu k_{l,-}}\right)}\left(\frac{L_\perp^4}{4!}+\frac{1}{2}L_\perp^2 \zeta_2+\frac{2}{3}L_\perp\zeta_3+\frac{7}{4}\zeta_4\right)\nonumber\\
&+\frac{L_\perp^5}{60}+\frac{1}{4}L_\perp^3\zeta_2+\frac{5}{12}L_\perp^2\zeta_3+\frac{7}{4}L_\perp \zeta_4+\frac{1}{2}\zeta_2\zeta_3+\frac{3}{10}\zeta_5\nonumber\\
&+\left(\frac{L_\perp^3}{12}+\frac{1}{2}L_\perp \zeta_2+\frac{\zeta_3}{3}\right)\mathrm{Li}_2\left(-\frac{k_{l,T}^2}{m_l^2}\right)+\left(\frac{L_\perp^2}{4}+\frac{\zeta_2}{2}\right)\mathrm{Li}_3\left(-\frac{k_{l,T}^2}{m_l^2}\right)\nonumber\\
&+\frac{1}{2}L_\perp\mathrm{Li}_4\left(-\frac{k_{l,T}^2}{m_l^2}\right)+\frac{1}{2}\mathrm{Li}_5\left(-\frac{k_{l,T}^2}{m_l^2}\right)\bigg]+\mathcal{O}(\alpha_1,\epsilon^4)\Bigg\}\,.\label{eq:E1l0mlavg}
\end{align}
The rapidity singularity is now expressed as $1/\alpha_1$. Similar to $\epsilon$, the expression inside the curly brackets in eq.~\eqref{eq:E1l0mlavg} has uniform transcendental weight $2$ if we assign a weight of $-1$ to $\alpha_1$. 

The $l$-independent eikonal integral, the azimuthally-averaged counterpart of eq.~\eqref{eq:TMDeikonalspecial}, can be directly deduced from $\bar{\mathcal{E}}_{\perp,1l}^{(0,m_l)}(b_{T},\mu)$ by replacing $k_l$ with $k_1+k_2$:
\begin{equation}
\bar{\mathcal{E}}_{\perp,1}(b_{T},\mu)=\left.\bar{\mathcal{E}}_{\perp,1l}^{(0,m_l)}(b_{T},\mu)\right|_{k_l\to k_1+k_2}=\left.\bar{\mathcal{E}}_{\perp,1l}^{(0,m_l)}(b_{T},\mu)\right|_{m_l\to M,k_{l,\pm}\to M,k_{l,T}\to 0}\,.\label{eq:EkbTavg}
\end{equation}
After combining eqs.~\eqref{eq:EkbTavg}, \eqref{eq:E1leikonal0ml1}, and \eqref{eq:E1l0mlavg}, the explicit expression for $\bar{\mathcal{E}}_{\perp,1}(b_{T},\mu)$ is:
\begin{equation}
\begin{aligned}
\bar{\mathcal{E}}_{\perp,1}(b_{T},\mu)=&\frac{(4\pi)^\epsilon}{\Gamma(1-\epsilon)}\Bigg\{\frac{1}{\alpha_1}\bigg[\frac{1}{\epsilon}+L_\perp+\epsilon\left(\frac{L_\perp^2}{2!}+\zeta_2\right)+\epsilon^2\left(\frac{L_\perp^3}{3!}+L_\perp \zeta_2+\frac{2}{3}\zeta_3\right)\\
&+\epsilon^3\left(\frac{L_\perp^4}{4!}+\frac{1}{2}L_\perp^2\zeta_2+\frac{2}{3}L_\perp \zeta_3+\frac{7}{4}\zeta_4\right)\\
&+\epsilon^4\left(\frac{L_\perp^5}{5!}+\frac{1}{6}L_\perp^3\zeta_2+\frac{1}{3}L_\perp^2\zeta_3+\frac{7}{4}L_\perp\zeta_4+\frac{2}{3}\zeta_2\zeta_3+\frac{2}{5}\zeta_5\right)+\mathcal{O}(\epsilon^5)\bigg]\\
&-\frac{1}{2\epsilon^2}+\frac{\log{\left(\frac{\nu_1}{\mu}\right)}}{\epsilon}+\log{\left(\frac{\nu_1}{\mu}\right)}L_\perp+\frac{L_\perp^2}{4}\\
&+\epsilon\bigg[\log{\left(\frac{\nu_1}{\mu}\right)}\left(\frac{L_\perp^2}{2!}+\zeta_2\right)+\frac{L_\perp^3}{6}+\frac{1}{2}L_\perp\zeta_2+\frac{\zeta_3}{6}\bigg]\\
&+\epsilon^2\bigg[\log{\left(\frac{\nu_1}{\mu}\right)}\left(\frac{L_\perp^3}{3!}+L_\perp \zeta_2+\frac{2}{3}\zeta_3\right)+\frac{L_\perp^4}{16}+\frac{1}{2}L_\perp^2 \zeta_2+\frac{1}{2}L_\perp \zeta_3+\frac{7}{8}\zeta_4\bigg]\\
&+\epsilon^3\bigg[\log{\left(\frac{\nu_1}{\mu}\right)}\left(\frac{L_\perp^4}{4!}+\frac{1}{2}L_\perp^2 \zeta_2+\frac{2}{3}L_\perp\zeta_3+\frac{7}{4}\zeta_4\right)\\
&+\frac{L_\perp^5}{60}+\frac{1}{4}L_\perp^3\zeta_2+\frac{5}{12}L_\perp^2\zeta_3+\frac{7}{4}L_\perp \zeta_4+\frac{1}{2}\zeta_2\zeta_3+\frac{3}{10}\zeta_5\bigg]+\mathcal{O}(\alpha_1,\epsilon^4)\Bigg\}\,.\label{eq:eikonal1avgexp}
\end{aligned}
\end{equation}

The final rapidity-divergence-subtracted one-massless-one-massive eikonal integral is:
\begin{equation}
\begin{aligned}
&\bar{\mathcal{E}}_{\perp,1l}^{(0,m_l)}(b_{T},\mu)-\bar{\mathcal{E}}_{\perp,1}(b_{T},\mu)\\
=&\frac{(4\pi)^\epsilon}{\Gamma(1-\epsilon)}\Bigg\{-\frac{\log{\left(\frac{k_{l,-}}{m_l}\right)}}{\epsilon}-L_\perp \log{\left(\frac{k_{l,-}}{m_l}\right)}+\frac{1}{2}\mathrm{Li}_2\left(-\frac{k_{l,T}^2}{m_l^2}\right)\\
 &+\epsilon\bigg[-\left(\frac{L_\perp^2}{2!}+\zeta_2\right)\log{\left(\frac{k_{l,-}}{m_l}\right)}+\frac{1}{2}L_\perp \mathrm{Li}_2\left(-\frac{k_{l,T}^2}{m_l^2}\right)+\frac{1}{2}\mathrm{Li}_3\left(-\frac{k_{l,T}^2}{m_l^2}\right)\bigg]\\
 &+\epsilon^2\bigg[-\log{\left(\frac{k_{l,-}}{m_l}\right)}\left(\frac{L_\perp^3}{3!}+L_\perp \zeta_2+\frac{2}{3}\zeta_3\right)+\left(\frac{L_\perp^2}{4}+\frac{\zeta_2}{2}\right)\mathrm{Li}_2\left(-\frac{k_{l,T}^2}{m_l^2}\right)\\
 &+\frac{1}{2}L_\perp \mathrm{Li}_3\left(-\frac{k_{l,T}^2}{m_l^2}\right)+\frac{1}{2}\mathrm{Li}_4\left(-\frac{k_{l,T}^2}{m_l^2}\right)\bigg]\\
&+\epsilon^3\bigg[-\log{\left(\frac{k_{l,-}}{m_{l}}\right)}\left(\frac{L_\perp^4}{4!}+\frac{1}{2}L_\perp^2 \zeta_2+\frac{2}{3}L_\perp\zeta_3+\frac{7}{4}\zeta_4\right)\\
&+\left(\frac{L_\perp^3}{12}+\frac{1}{2}L_\perp \zeta_2+\frac{\zeta_3}{3}\right)\mathrm{Li}_2\left(-\frac{k_{l,T}^2}{m_l^2}\right)+\left(\frac{L_\perp^2}{4}+\frac{\zeta_2}{2}\right)\mathrm{Li}_3\left(-\frac{k_{l,T}^2}{m_l^2}\right)\\
&+\frac{1}{2}L_\perp\mathrm{Li}_4\left(-\frac{k_{l,T}^2}{m_l^2}\right)+\frac{1}{2}\mathrm{Li}_5\left(-\frac{k_{l,T}^2}{m_l^2}\right)\bigg]+\mathcal{O}(\epsilon^4)\Bigg\}\,.\label{eq:eikonalmasslessmassiveavg}
\end{aligned}
\end{equation}
Equation \eqref{eq:eikonalmasslessmassiveavg} agrees with eq.~(96) in ref.~\cite{Catani:2023tby} up to $\mathcal{O}(\epsilon^1)$, modulo the dependence on an overall factor $8\pi^2\mu^{2\epsilon}/\left(2\pi\right)^{d-1}$.
The counterpart for the second beam can be easily determined based on the relation in eq.~\eqref{eq:E2bTmE1bTazimuthavg}.

\subsubsection{Azimuthal-angle-dependent result}

For the full azimuthal-angle-dependent result, we need to perform the Laurent series expansion of the two infinitesimal variables $\alpha_1$ and $\epsilon$ according to eq.~\eqref{eq:order4regulators}. On the right-hand side of eq.~\eqref{eq:E1l0ml}, there are three different Gaussian hypergeometric functions  that need to be addressed. For the first hypergeometric function in the real part, we can directly apply the series representation in eq.~\eqref{eq:series42F1}. Up to $\mathcal{O}(\alpha_1)$, this function can be expanded with the help of {\tt HypExp} as follows:
\begin{eqnarray}
&&{}_2\hspace{-0.7mm}F_1\left(-\frac{\alpha_1}{2},-\frac{\alpha_1}{2}-\epsilon; \frac{1}{2}; x\right)\nonumber\\
&=&1-\frac{\alpha_1}{2}\sum_{i=1}^{+\infty}{\frac{1}{i}\frac{(1)_i(-\epsilon)_i}{(\frac{1}{2})_i}\frac{x^i}{i!}}+\mathcal{O}(\alpha_1^2)\nonumber\\
&=&1+\frac{\alpha_1}{2}\left[2\epsilon x\,{}_3\hspace{-0.7mm}F_2\left(1,1,1-\epsilon; \frac{3}{2}, 2; x\right)\right]+\mathcal{O}(\alpha_1^2)\nonumber\\
&=&1+\alpha_1\Bigg\{-\frac{\epsilon}{4}\log^2{\left(\frac{1-y}{1+y}\right)}+\epsilon^2\Bigg[2\left(\mathrm{Li}_3\left(\frac{1+y}{2}\right)+\mathrm{Li}_3\left(\frac{1-y}{2}\right)\right)\nonumber\\
&&+\frac{1}{2}\left(\mathrm{Li}_2\left(\frac{1+y}{2}\right)-\mathrm{Li}_2\left(\frac{1-y}{2}\right)\right)\log{\left(\frac{1-y}{1+y}\right)}\nonumber\\
&&+\frac{1}{12}\log^3{\left(\frac{1-y^2}{4}\right)}-\zeta_2\log{\left(\frac{1-y^2}{4}\right)}-\frac{7}{2}\zeta_3\Bigg]\nonumber\\
&&+\epsilon^3\Bigg[\left(\mathrm{Li}_3\left(\frac{1+y}{2}\right)-\mathrm{Li}_3\left(\frac{1-y}{2}\right)\right)\log{\left(\frac{1-y}{1+y}\right)}\nonumber\\
&&+\frac{1}{4}\left(\mathrm{Li}_2\left(\frac{1+y}{2}\right)-\mathrm{Li}_2\left(\frac{1-y}{2}\right)\right)^2+\frac{\zeta_2}{2}\log^2{\left(\frac{1-y}{1+y}\right)}\nonumber\\
&&+\frac{1}{4}\left(\mathrm{Li}_2\left(\frac{1+y}{2}\right)-\mathrm{Li}_2\left(\frac{1-y}{2}\right)\right)\left(\log^2{\left(\frac{1+y}{2}\right)}-\log^2{\left(\frac{1-y}{2}\right)}\right)\nonumber\\
&&-\frac{1}{48}\log^2{\left(\frac{1-y}{1+y}\right)}\left(\log^2{\left(\frac{1-y}{1+y}\right)}+12\log{\left(\frac{1-y}{2}\right)}\log{\left(\frac{1+y}{2}\right)}\right)\Bigg]\nonumber\\
&&+\mathcal{O}(\epsilon^4)\Bigg\}+\mathcal{O}(\alpha_1^2)\,,\label{eq:hypergeometric1F1exp10}
\end{eqnarray}
In the last identity, we have set $x=y^2/(y^2-1)$ and $y=\sqrt{x/(x-1)}$ (cf. eq.~\eqref{eq:ydefinition}), while imposing the constraints $x<0$ and $0\leq y<1$. 

Applying the same approach to the other two hypergeometric functions in the imaginary part of eq.~\eqref{eq:E1l0ml} is challenging due to the difficult series summation in eq.~\eqref{eq:series42F1} after expanding over $\alpha_1$. Instead, we use the integral representation of the ordinary hypergeometric function:
\begin{equation}
\begin{aligned}
{}_2\hspace{-0.7mm}F_1\left(a,b; c; x\right)=&\frac{\Gamma(c)}{\Gamma(b)\Gamma(c-b)}\int_0^1{dt t^{b-1}(1-t)^{c-b-1}(1-tx)^{-a}}\,,\\ 
& \quad \quad \quad\quad \mathrm{when} \quad 0<\Re{(b)}<\Re{(c)}, \;|\arg{\left(1-x\right)}|<\pi\,.\label{eq:integralrep42F1}
\end{aligned}
\end{equation}
By setting $a=-\alpha_1/2, b=-\alpha_1/2-\epsilon$, and $c=1/2$ and keeping the series expansion of the integrand in eq.~\eqref{eq:integralrep42F1} up to $\mathcal{O}(\alpha_1)$, the second equality of eq.~\eqref{eq:hypergeometric1F1exp10} can be directly reproduced by imposing the constraints $-\frac{1}{2}<\Re{(\epsilon)}<0$ and $x<0$. Additionally, the recurrence relation
\begin{equation}
{}_2\hspace{-0.7mm}F_1\left(a,b; -\frac{1}{2}; x\right)={}_2\hspace{-0.7mm}F_1\left(a,b; \frac{1}{2}; x\right)-4abx\,{}_2\hspace{-0.7mm}F_1\left(1+a,1+b; \frac{3}{2}; x\right)
\end{equation}
allows us to express the function ${}_2\hspace{-0.7mm}F_1\left((1-\alpha_1)/2,(1-\alpha_1-2\epsilon)/2; -1/2; x\right)$ in terms of the existing hypergeometric function ${}_2\hspace{-0.7mm}F_1\left((1-\alpha_1)/2,(1-\alpha_1-2\epsilon)/2; 1/2; x\right)$ and a new function
${}_2\hspace{-0.7mm}F_1\left((3-\alpha_1)/2,(3-\alpha_1-2\epsilon)/2; 3/2; x\right)$. This conversion avoids a double pole in the integral representation eq.~\eqref{eq:integralrep42F1} at $t=1$. For convenience, we again perform a change of variables:
\begin{equation}
y\equiv \sqrt{\frac{x}{x-1}}\quad \Longrightarrow \quad x=\frac{y^2}{y^2-1}\,.
\end{equation}
The leading terms in $\alpha_1$ can then be obtained at all orders in $\epsilon$:
\begin{eqnarray}
\lim_{\alpha_1\to0}{}_2\hspace{-0.7mm}F_1\left(\frac{1-\alpha_1}{2},\frac{1-\alpha_1}{2}-\epsilon; \frac{1}{2}; x\right)&=&{}_2\hspace{-0.7mm}F_1\left(\frac{1}{2},\frac{1}{2}-\epsilon; \frac{1}{2}; x\right)=(1-x)^{-\frac{1}{2}+\epsilon}=\left(1-y^2\right)^{\frac{1}{2}-\epsilon}\,,\nonumber\\ \\
\lim_{\alpha_1\to0}{}_2\hspace{-0.7mm}F_1\left(\frac{3-\alpha_1}{2},\frac{3-\alpha_1}{2}-\epsilon; \frac{3}{2}; x\right)&=&{}_2\hspace{-0.7mm}F_1\left(\frac{3}{2},\frac{3}{2}-\epsilon; \frac{3}{2}; x\right)=(1-x)^{-\frac{3}{2}+\epsilon}=\left(1-y^2\right)^{\frac{3}{2}-\epsilon}\,.\nonumber\\
\end{eqnarray}
For the $\mathcal{O}(\alpha_1)$ terms, we keep the $\mathcal{O}(\alpha_1)$ series expansion in the integrand of eq.~\eqref{eq:integralrep42F1}. These terms can generally be written in the form
\begin{equation}
\alpha_1 \int_0^1{dt (1-t)^{-1+\epsilon}f(t,x,\epsilon)}\,,
\end{equation}
where $f(t,x,\epsilon)$ is determined from the integrand in eq.~\eqref{eq:integralrep42F1} after substituting the specific values of $a,b$, and $c$. Since there remains single poles at $t=1$, we decompose the integral into two parts:
\begin{equation}
\begin{aligned}
\alpha_1 \int_0^1{dt (1-t)^{-1+\epsilon}f(t,x,\epsilon)}=&\alpha_1 \int_0^1{dt (1-t)^{-1+\epsilon}f(1,x,\epsilon)}\\
&+\alpha_1 \int_0^1{dt (1-t)^{-1+\epsilon}\left[f(t,x,\epsilon)-f(1,x,\epsilon)\right]}\,.\label{eq:hypergeometricOa10}
\end{aligned}
\end{equation}
The first term on the right-hand side is straightforward and can be integrated at all orders in $\epsilon$, while the second term, which is regular at $t=1$, is more complex and deserves further explanation. To evaluate the second term, we express it in terms of Goncharov's generalization of polylogarithms (GPLs)~\cite{Goncharov:1998,Goncharov:2001}, which are well-studied mathematical functions defined iteratively as
\begin{equation}
G(a_1,a_2,\ldots,a_n; z)=\int_0^{z}{\frac{dt}{t-a_1}G(a_2,\ldots,a_n; t)}\label{eq:GPLdef}
\end{equation}
with $G(;z)=1$ and $G(\underbrace{0,\ldots,0}_{n};z)=\log^n{\left(z\right)}/n!$. The indices $a_i$ and the argument $z$ are generally complex variables. Like other multiple polylogarithms, $G(a_1,a_2,\ldots,a_n; z)$ is a transcendental function with weight $n$. Efficient numerical algorithms~\cite{Vollinga:2004sn} for evaluating GPLs have been implemented in several public tools, including {\tt GiNaC}~\cite{Bauer:2000cp},~\footnote{The {\tt GiNaC} implementation for evaluating GPLs can also be accessed through the {\sc\small Mathematica} package {\tt PolyLogTools}~\cite{Duhr:2019tlz}.} {\tt handyG}~\cite{Naterop:2019xaf}, and {\tt FastGPL}~\cite{Wang:2021imw}.

After performing the series expansion of the infinitesimal regulator $\epsilon$ in the integrand of the second term, two square roots, $\sqrt{t}$ and $\sqrt{1-tx}$, appear. To rationalize them simultaneously, we change the integration variable to $z\equiv \sqrt{-tx/(1-tx)}$ (i.e., $t=-z^2/(1-z^2)/x$). Since the argument $x$ in the hypergeometric functions of eq.~\eqref{eq:E1l0ml} is strictly negative, we take the integration range of $z$ from $z=0$ to $z=y$ with $0\leq y<1$. At each order in $\epsilon$,  the second term in eq.~\eqref{eq:hypergeometricOa10} can be expressed in terms of GPLs with the possible indices $\left\{0,1,-1,y,-y\right\}$ and integration kernels of the form $\left\{1/(z-1),1/(z+1),1/(z-y),1/(z+y)\right\}$, following the iterative definition of GPLs in eq.~\eqref{eq:GPLdef}. During intermediate steps, we encounter singular GPLs of the form
\begin{equation}
G(\underbrace{y,\ldots,y}_{n}, a_{n+1},\ldots,a_{m};y), \quad n\geq 1,\; m\geq n,\; a_{n+1}\neq y\,.
\end{equation}
To handle these terms, we convert them into sums of products of pure singular GPLs $G(y,\ldots,y;y)$, and regular GPLs, $G(a_1,\ldots,a_k;y)$ with $a_1\neq y$, using the shuffle algebra. The singular GPLs $G(y,\ldots,y;y)$ cancel out in the final expressions. 

In the case where $a=(3-\alpha_1)/2,b=(3-\alpha_1-2\epsilon)/2$, and $c=3/2$, we additionally need to handle the following integral:
\begin{equation}
\int_0^y{dz G(a_1,a_{2},\ldots,a_n;z)}\,,
\end{equation}
which can be solved recursively using integration by parts:
\begin{equation}
\int_0^y{dz G(a_1,a_{2},\ldots,a_n;z)}=(y-a_1)G(a_1,a_{2},\ldots,a_n;y)-\int_0^y{dz G(a_{2},\ldots,a_n;z)}\,.
\end{equation}
In this way, the two hypergeometric functions, ${}_2\hspace{-0.7mm}F_1\left((1-\alpha_1)/2,(1-\alpha_1-2\epsilon)/2; 1/2; x\right)$ and ${}_2\hspace{-0.7mm}F_1\left((3-\alpha_1)/2,(3-\alpha_1-2\epsilon)/2; 3/2; x\right)$, can be expressed in terms of GPLs up to $\mathcal{O}(\alpha_1)$ and to all orders in the dimensional regulator $\epsilon$. We observe that, at each order in $\epsilon$, the expansion coefficients of ${}_2\hspace{-0.7mm}F_1\left((1-\alpha_1)/2,(1-\alpha_1-2\epsilon)/2; 1/2; x\right)$ exhibit the UT weight property. In contrast, this property does not hold for the $\epsilon$ expansion of \\
${}_2\hspace{-0.7mm}F_1\left((3-\alpha_1)/2,(3-\alpha_1-2\epsilon)/2; 3/2; x\right)$. Nevertheless, the final eikonal integral still maintains uniform transcendentality. The full azimuthal-angle-dependent TMD eikonal integral is then given by
\begin{equation}
\begin{aligned}
\bar{\mathcal{E}}_{\perp,1l}^{(0,m_l)}(\harpoon{b}_{T},\mu)=&\frac{(4\pi)^\epsilon}{\Gamma(1-\epsilon)}\Bigg\{\frac{1}{\alpha_1}\bigg[\frac{1}{\epsilon}+L_\perp+\epsilon\left(\frac{L_\perp^2}{2!}+\zeta_2\right)+\epsilon^2\left(\frac{L_\perp^3}{3!}+L_\perp \zeta_2+\frac{2}{3}\zeta_3\right)\\
&+\epsilon^3\left(\frac{L_\perp^4}{4!}+\frac{1}{2}L_\perp^2\zeta_2+\frac{2}{3}L_\perp \zeta_3+\frac{7}{4}\zeta_4\right)\\
&+\epsilon^4\left(\frac{L_\perp^5}{5!}+\frac{1}{6}L_\perp^3\zeta_2+\frac{1}{3}L_\perp^2\zeta_3+\frac{7}{4}L_\perp\zeta_4+\frac{2}{3}\zeta_2\zeta_3+\frac{2}{5}\zeta_5\right)+\mathcal{O}(\epsilon^5)\bigg]\\
&-\frac{1}{2\epsilon^2}+\frac{\log{\left(\frac{\nu_1 m_l}{\mu k_{l,-}}\right)}}{\epsilon}+\log{\left(\frac{\nu_1 m_l}{\mu k_{l,-}}\right)} L_\perp+\frac{L_\perp^2}{4}+V_0\\
&+\epsilon\bigg[\log{\left(\frac{\nu_1 m_l}{\mu k_{l,-}}\right)}\left(\frac{L_\perp^2}{2!}+\zeta_2\right)+\frac{L_\perp^3}{6}+\frac{1}{2}L_\perp \zeta_2+\frac{1}{6}\zeta_3+V_1\bigg]\\
&+\epsilon^2\bigg[\log{\left(\frac{\nu_1 m_l}{\mu k_{l,-}}\right)}\left(\frac{L_\perp^3}{3!}+L_\perp\zeta_2+\frac{2}{3}\zeta_3\right)+\frac{L_\perp^4}{16}+\frac{1}{2}L_\perp^2\zeta_2+\frac{1}{2}L_\perp\zeta_3+\frac{7}{8}\zeta_4+V_2\bigg]\\
&+\epsilon^3\bigg[\log{\left(\frac{\nu_1 m_l}{\mu k_{l,-}}\right)}\left(\frac{L_\perp^4}{4!}+\frac{1}{2}L_\perp^2\zeta_2+\frac{2}{3}L_\perp\zeta_3+\frac{7}{4}\zeta_4\right)\\
&+\frac{L_\perp^5}{60}+\frac{1}{4}L_\perp^3 \zeta_2+\frac{5}{12}L_\perp^2\zeta_3+\frac{7}{4}L_\perp \zeta_4+\frac{1}{2}\zeta_2\zeta_3+\frac{3}{10}\zeta_5+V_3\bigg]+\mathcal{O}(\alpha_1,\epsilon^4)\Bigg\}\,,\label{eq:E1l0mlexp}
\end{aligned}
\end{equation}
where the expressions for $V_i$ are given by
\begin{align}
V_0=&\frac{1}{2} \left[-G(-1,-1;y)+G(-1,1;y)+G(1,-1;y)-G(1,1;y)\right]\nonumber\\
&+i\pi \frac{\harpoon{k}_{l,T}\cdot \harpoon{b}_{T}}{|\harpoon{k}_{l,T}\cdot \harpoon{b}_{T}|} \frac{1}{2} \left[G(1;y)-G(-1;y)\right]\,,
\end{align}
and
\begin{align}
V_1=&\frac{1}{2} L_{\perp}\left(-G(-1,-1;y)+G(-1,1;y)+G(1,-1;y)-G(1,1;y)\right)\nonumber\\
&+\frac{1}{2}\Bigg(G(-1,-1,-1;y)-G(-1,-1,1;y)+G(-1,1,-1;y)-G(-1,1,1;y)\nonumber\\
&-G(1,-1,-1;y)+G(1,-1,1;y)-G(1,1,-1;y)+G(1,1,1;y)\Bigg)\nonumber\\
&+i\pi\frac{\harpoon{k}_{l,T}\cdot \harpoon{b}_{T}}{|\harpoon{k}_{l,T}\cdot \harpoon{b}_{T}|}\Bigg[\frac{1}{2} \bigg(y G(-1,-1;y)-y G(-1,1;y)+(1+y) G(-1,-y;y)\nonumber\\
&+(1-y)G(-1,y;y)+(1+y) G(0,1;y)-y G(1,-1;y)-G(1,1;y)+(y-1)G(1,-y;y)\nonumber\\
&+2 G(-y,-1;y)-2 G(-y,1;y)\bigg)-\frac{1}{2} \left(G(-1;y)-G(1;y)\right)\left(L_{\perp}+2 \log (2)\right)\Bigg]\,,
\end{align}
and
\begin{align}
V_2=&\frac{1}{4} \left(-G(-1,-1;y)+G(-1,1;y)+G(1,-1;y)-G(1,1;y)\right) \left(2 \zeta_2+L_{\perp}^2\right)\nonumber\\
&+\frac{1}{2} L_{\perp}\bigg(G(-1,-1,-1;y)-G(-1,-1,1;y)+G(-1,1,-1;y)-G(-1,1,1;y)\nonumber\\
&-G(1,-1,-1;y)+G(1,-1,1;y)-G(1,1,-1;y)+G(1,1,1;y)\bigg)\nonumber\\
&+\frac{1}{2}\bigg(-G(-1,-1,-1,-1;y)+G(-1,-1,-1,1;y)-G(-1,-1,1,-1;y)\nonumber\\
&+G(-1,-1,1,1;y)-G(-1,1,-1,-1;y)+G(-1,1,-1,1;y)-G(-1,1,1,-1;y)\nonumber\\
&+G(-1,1,1,1;y)+G(1,-1,-1,-1;y)-G(1,-1,-1,1;y)+G(1,-1,1,-1;y)\nonumber\\
&-G(1,-1,1,1;y)+G(1,1,-1,-1;y)-G(1,1,-1,1;y)+G(1,1,1,-1;y)-G(1,1,1,1;y)\bigg)\nonumber\\
&+i\pi\frac{\harpoon{k}_{l,T}\cdot \harpoon{b}_{T}}{|\harpoon{k}_{l,T}\cdot \harpoon{b}_{T}|}\Bigg[\frac{1}{2} L_{\perp} \bigg(y G(-1,-1;y)-y G(-1,1;y)+yG(-1,-y;y)+G(-1,-y;y)\nonumber\\
&-y G(-1,y;y)+G(-1,y;y)+y G(0,1;y)+G(0,1;y)-yG(1,-1;y)\nonumber\\
&-G(1,1;y)+y G(1,-y;y)-G(1,-y;y)+2 G(-y,-1;y)-2 G(-y,1;y)\bigg)\nonumber\\
&+\log (2) \bigg(yG(-1,-1;y)-y G(-1,1;y)+y G(-1,-y;y)+G(-1,-y;y)-y G(-1,y;y)\nonumber\\
&+G(-1,y;y)+yG(0,1;y)+G(0,1;y)-y G(1,-1;y)-G(1,1;y)\nonumber\\
&+y G(1,-y;y)-G(1,-y;y)+2 G(-y,-1;y)-2G(-y,1;y)\bigg)\nonumber\\
&+\frac{1}{2} \bigg(-(2 y+1) G(-1,-1,-1;y)-(2 y+1) G(-1,-1,1;y)+3 yG(-1,-1,-y;y)\nonumber\\
&+2 G(-1,-1,-y;y)+y G(-1,-1,y;y)+y G(-1,0,1;y)+4 y G(-1,0,-y;y)\nonumber\\
&+2y G(-1,0,y;y)-2 y G(-1,1,-1;y)-G(-1,1,-1;y)+5 y G(-1,1,1;y)\nonumber\\
&-G(-1,1,1;y)-2 yG(-1,1,-y;y)+2 G(-1,1,-y;y)-y G(-1,1,y;y)\nonumber\\
&+3 y G(-1,-y,-1;y)+2 G(-1,-y,-1;y)-3y G(-1,-y,1;y)+2 G(-1,-y,1;y)\nonumber\\
&-3 y G(-1,-y,-y;y)-4 G(-1,-y,-y;y)+3 yG(-1,-y,y;y)-2 G(-1,-y,y;y)\nonumber\\
&-y G(-1,y,-y;y)-2 G(-1,y,-y;y)+y G(-1,y,y;y)+yG(0,-1,1;y)\nonumber\\
&+4 y G(0,-1,-y;y)+2 y G(0,-1,y;y)+y G(0,1,-1;y)-5 y G(0,1,1;y)\nonumber\\
&+4 yG(0,1,-y;y)-2 G(0,1,-y;y)+2 y G(0,1,y;y)+6 y G(0,-y,-1;y)\nonumber\\
&+6 y G(0,-y,1;y)-2G(0,-y,1;y)+2 y G(1,-1,-1;y)+G(1,-1,-1;y)\nonumber\\
&+y G(1,-1,1;y)+G(1,-1,1;y)-2 yG(1,-1,-y;y)-2 G(1,-1,-y;y)\nonumber\\
&-y G(1,-1,y;y)-3 y G(1,0,1;y)+4 y G(1,0,-y;y)+2 yG(1,0,y;y)\nonumber\\
&+y G(1,1,-1;y)+G(1,1,-1;y)+2 y G(1,1,1;y)+G(1,1,1;y)\nonumber\\
&+y G(1,1,-y;y)-2y G(1,1,y;y)-3 y G(1,-y,-1;y)-2 G(1,-y,-1;y)\nonumber\\
&+3 y G(1,-y,1;y)-3 yG(1,-y,-y;y)+4 G(1,-y,-y;y)+3 y G(1,-y,y;y)\nonumber\\
&-y G(1,y,-1;y)-y G(1,y,1;y)-yG(1,y,-y;y)+y G(1,y,y;y)\nonumber\\
&+4 y G(-y,-1,-1;y)+2 G(-y,-1,-1;y)-4 y G(-y,-1,1;y)+2G(-y,-1,1;y)\nonumber\\
&-4 y G(-y,-1,-y;y)-6 G(-y,-1,-y;y)+4 y G(-y,-1,y;y)-2G(-y,-1,y;y)\nonumber\\
&-2 G(-y,0,1;y)-4 y G(-y,1,-1;y)-2 G(-y,1,-1;y)+4 y G(-y,1,1;y)\nonumber\\
&-4 yG(-y,1,-y;y)+6 G(-y,1,-y;y)+4 y G(-y,1,y;y)-6 y G(-y,-y,-1;y)\nonumber\\
&-8G(-y,-y,-1;y)-6 y G(-y,-y,1;y)+8 G(-y,-y,1;y)+6 y G(-y,y,-1;y)\nonumber\\
&+6 yG(-y,y,1;y)\bigg)-\frac{1}{4} (G(-1;y)-G(1;y)) \left(L_{\perp}^2+4 \log (2) L_{\perp}+4\zeta_2+4 \log ^2{(2)}\right)\Bigg]\,.
\end{align}
The expression of $V_3$ is too lengthy to be explicitly written here. We provide it along with those for $V_0,V_1$, and $V_2$ in the {\sc\small Mathematica} ancillary file\\
{\tt OneMasslessOneMassiveEikonal\_FullAzimuth.wl}. Overall, there are $6,20,86$, and $462$ GPLs involved in $V_0,V_1,V_2$, and $V_3$, respectively. We observe that although one hypergeometric function and some prefactors in eq.~\eqref{eq:E1l0ml} have different transcendental weights, the final TMD eikonal integral is ultimately UT. Our approach has the potential to be generalized for performing series expansions in more-than-one infinitesimal parameters (e.g., $\alpha_1$ and $\epsilon$ here) within hypergeometric functions, which goes beyond the capabilities of existing public tools. 

We have performed several numerical checks to validate eq.~\eqref{eq:E1l0mlexp}. Specifically, we compare our analytic result with the $\alpha_1$ and $\epsilon$ series expansions of the hypergeometric functions using the built-in {\sc\small Mathematica} function {\tt Series}, which allows us to evaluate their numerical values in the last equality of eq.~\eqref{eq:E1l0ml}. We randomly sample over 10k phase space points, and in all cases, we find perfect agreement. For these checks, we use {\tt handyG} to numerically evaluate the GPLs appearing in $V_i$.

In order to determine $\bar{\mathcal{E}}_{\perp,1}(\harpoon{b}_{T},\mu)$, we use the relation
\begin{eqnarray}
\bar{\mathcal{E}}_{\perp,1}(\harpoon{b}_{T},\mu)&=&\left.\bar{\mathcal{E}}_{\perp,1l}^{(0,m_l)}(\harpoon{b}_{T},\mu)\right|_{k_l\to k_1+k_2}=\left.\bar{\mathcal{E}}_{\perp,1l}^{(0,m_l)}(\harpoon{b}_{T},\mu)\right|_{m_l\to M,k_{l,\pm}\to M,\harpoon{k}_{l,T}\to \harpoon{0}}\,,\label{eq:EkbTfullazimu}
\end{eqnarray}
Since $V_i$ vanishes when $y=0$ or $x=0$, combining eqs.~\eqref{eq:EkbTfullazimu}, \eqref{eq:E1l0ml}, and \eqref{eq:E1l0mlexp} gives
\begin{equation}
\begin{aligned}
\bar{\mathcal{E}}_{\perp,1}(\harpoon{b}_{T},\mu)=&\bar{\mathcal{E}}_{\perp,1}(b_{T},\mu)\,,
\end{aligned}
\end{equation}
where the right-hand side is given in eq.~\eqref{eq:eikonal1avgexp}.
The rapidity-divergence-subtracted eikonal integral, retaining full azimuthal-angle dependence, is
\begin{equation}
\begin{aligned}
 &\bar{\mathcal{E}}_{\perp,1l}^{(0,m_l)}(\harpoon{b}_{T},\mu)-\bar{\mathcal{E}}_{\perp,1}(\harpoon{b}_{T},\mu)\\
 =&\frac{(4\pi)^\epsilon}{\Gamma(1-\epsilon)}\Bigg\{-\frac{\log{\left(\frac{k_{l,-}}{m_l}\right)}}{\epsilon}-L_\perp \log{\left(\frac{k_{l,-}}{m_l}\right)}+V_0\\
 &+\epsilon\bigg[-\left(\frac{L_\perp^2}{2!}+\zeta_2\right)\log{\left(\frac{k_{l,-}}{m_l}\right)}+V_1\bigg]\\
 &+\epsilon^2\bigg[-\log{\left(\frac{k_{l,-}}{m_l}\right)}\left(\frac{L_\perp^3}{3!}+L_\perp \zeta_2+\frac{2}{3}\zeta_3\right)+V_2\bigg]\\
&+\epsilon^3\bigg[-\log{\left(\frac{k_{l,-}}{m_{l}}\right)}\left(\frac{L_\perp^4}{4!}+\frac{1}{2}L_\perp^2 \zeta_2+\frac{2}{3}L_\perp\zeta_3+\frac{7}{4}\zeta_4\right)+V_3\bigg]\\
&+\mathcal{O}(\epsilon^4)\Bigg\}\,.\label{eq:eikonalmasslessmassiveazimu}
\end{aligned}
\end{equation}
For $k=2$, the rapidity-divergence-subtracted integral is given by eq.~\eqref{eq:eikonalmasslessmassiveazimu}, replacing $k_{l,-}$ with $k_{l,+}$ according to eq.~\eqref{eq:E2bTmE1bTfullazimuth}.

\subsection{Two massive case\label{sec:TMD1Leikonaltwomassive}}

Finally, we turn to what is likely the most complex case in this section: the two-massive-particle scenario. Specifically, we consider $k\neq l$ with $k,l\geq 3$ and $m_k,m_l\neq 0$. For simplicity, we introduce the velocity-like variables
\begin{equation}
v_k^\mu\equiv \frac{k_k^\mu}{m_k},\quad v_l^\mu\equiv \frac{k_l^\mu}{m_l}\,,
\end{equation}
such that $v_k^2=v_l^2=1$ and $v_k\cdot v_l\geq 1$. The corresponding transverse components are denoted as $\harpoon{v}_{k,T}$ and $\harpoon{v}_{l,T}$, respectively. This two-massive case can be mapped onto the massive self-eikonal case discussed in section \ref{sec:TMD1Leikonalselfmassive} by employing Feynman parametrization:
\begin{equation}
\frac{1}{\left(v_k\cdot k_g\right)\left(v_l\cdot k_g\right)}=\int_0^1{\frac{du}{\left[uv_k\cdot k_g+(1-u)v_l\cdot k_g\right]^2}}\,.\label{eq:Feynmanparameter}
\end{equation}
Since there is no rapidity divergence, we can safely set $\alpha_1=0$ before performing any integration.

\subsubsection{Azimuthal-angle-averaged result}

Using eq.~\eqref{eq:Feynmanparameter}, the azimuthal-angle-averaged TMD eikonal integral can be expressed as
\begin{eqnarray}
\bar{\mathcal{E}}_{\perp,kl}^{(m_k,m_l)}(b_{T},\mu)&=&\pi^\epsilon \left(b_T\mu\right)^{2\epsilon}\Gamma\left(-\epsilon\right)\underbrace{\int_0^1{du \frac{v_k\cdot v_l}{u_{kl}^2}{}_2\hspace{-0.7mm}F_1\left(1,-\epsilon;1-\epsilon;-\frac{u_{kl,T}^2}{u_{kl}^2}\right)}}_{\equiv v_k\cdot v_l I_{kl}(\epsilon)}\nonumber\\
&=&\frac{(4\pi)^\epsilon}{\Gamma(1-\epsilon)}v_k\cdot v_l\Bigg[-\frac{I_{kl}^{(0)}}{\epsilon}-L_\perp I_{kl}^{(0)}-I_{kl}^{(1)}-\epsilon\left(\left(\frac{L_\perp^2}{2!}+\zeta_2\right) I_{kl}^{(0)}+L_\perp I_{kl}^{(1)}+I_{kl}^{(2)}\right)\nonumber\\
&&-\epsilon^2\left(\left(\frac{L_\perp^3}{3!}+L_\perp \zeta_2+\frac{2}{3}\zeta_3\right)I_{kl}^{(0)}+\left(\frac{L_\perp^2}{2!}+\zeta_2\right)I_{kl}^{(1)}+L_\perp I_{kl}^{(2)}+I_{kl}^{(3)}\right)\nonumber\\
&&-\epsilon^3\bigg(\left(\frac{L_\perp^4}{4!}+\frac{1}{2}L_\perp^2\zeta_2+\frac{2}{3}L_\perp\zeta_3+\frac{7}{4}\zeta_4\right)I_{kl}^{(0)}+\left(\frac{L_\perp^3}{3!}+L_\perp \zeta_2+\frac{2}{3}\zeta_3\right)I_{kl}^{(1)}\nonumber\\
&&+\left(\frac{L_\perp^2}{2!}+\zeta_2\right)I_{kl}^{(2)}+L_\perp I_{kl}^{(3)}+I_{kl}^{(4)}\bigg)\Bigg]+\mathcal{O}(\epsilon^4)\,,\label{eq:twomassiveoneloopbTavgsoft}
\end{eqnarray}
where
\begin{equation}
u_{kl,T}^2=u^2v_{k,T}^2+(1-u)^2v_{l,T}^2+2u(1-u)\harpoon{v}_{k,T}\cdot \harpoon{v}_{l,T}\,.
\end{equation}
Using {\tt HypExp}~\cite{Huber:2007dx}, the hypergeometric function in the integrand can be expanded in $\epsilon$:
\begin{equation}
 {}_2\hspace{-0.7mm}F_1\left(1,-\epsilon;1-\epsilon;x\right)=1-\sum_{i=1}^{+\infty}{\epsilon^i\mathrm{Li}_i(x)}\,.
\end{equation}
The integral over the hypergeometric function, $I_{kl}(\epsilon)$, admits the Taylor expansion:
\begin{equation}
I_{kl}(\epsilon)=\sum_{i=0}^{+\infty}{\epsilon^i I_{kl}^{(i)}}\,,
\end{equation}
where the expansion coefficients are given by
\begin{eqnarray}
I_{kl}^{(0)}&=&\int_0^{1}{du\frac{1}{u_{kl}^2}}\,,\quad
I_{kl}^{(i\geq 1)}=-\int_0^1{du\frac{\mathrm{Li}_i\left(-\frac{u_{kl,T}^2}{u_{kl}^2}\right)}{u_{kl}^2}}\,.
\label{eq:Iklonemasslessonemassive}
\end{eqnarray}

The remaining one-fold integral over $u$ can be straightforwardly solved for $i=0$, yielding
\begin{eqnarray}
I_{kl}^{(0)}&=&\frac{1}{\sqrt{v_{kl}^{(+)}v_{kl}^{(-)}}}\log{\left(\frac{\sqrt{v_{kl}^{(+)}}+\sqrt{v_{kl}^{(-)}}}{\sqrt{v_{kl}^{(+)}}-\sqrt{v_{kl}^{(-)}}}\right)}\nonumber\\
&=&\frac{1}{v_k\cdot v_l}\frac{1}{2v_{kl} }\log{\left(\frac{1+v_{kl}}{1-v_{kl}}\right)}\,,\label{eq:Ikl0exp1}
\end{eqnarray}
where we have introduced the shorthand notations
\begin{eqnarray}
v_{kl}^{(\pm)}&\equiv& v_k\cdot v_l\pm 1\,,\label{eq:defvklpm}\\
v_{kl}&\equiv&\sqrt{1-\left(\frac{m_km_l}{k_k\cdot k_l}\right)^2}=\sqrt{1-\frac{1}{\left(v_k\cdot v_l\right)^2}}=\frac{\sqrt{v_{kl}^{(+)}v_{kl}^{(-)}}}{v_k\cdot v_l}\,.\label{eq:defvkl}
\end{eqnarray}
The definition of $v_{kl}$ follows the same convention as eq.(A.14) in ref.~\cite{Frederix:2009yq}.

For the general $i\geq 1$ case, we first express $\mathrm{Li}_i(x)$ in terms of GPLs:
\begin{equation}
\mathrm{Li}_i(x)=-G(\underbrace{0,\ldots,0}_{i-1},x^{-1};1)=-G(\underbrace{0,\ldots,0}_{i-1},1;x)\,.
\end{equation}
Thus, we obtain the Chen's iterated path integral representation~\cite{Chen:1977oja} for $I_{kl}^{(i)}$ with $i\geq 1$ as follows:
\begin{equation}
\begin{aligned}
I_{kl}^{(i)}=&-\int_0^1{du \frac{\mathrm{Li}_i\left(-\frac{u_{kl,T}^2}{u_{kl}^2}\right)}{u_{kl}^2}}=\int_0^1{du \frac{G(\underbrace{0,\ldots,0}_{i-1},1;-\frac{u_{kl,T}^2}{u_{kl}^2})}{u_{kl}^2}}\\
=&\frac{1}{2\sqrt{v_{kl}^{(+)}v_{kl}^{(-)}}}\int_0^1{\left(d\log{l_1(u)}-d\log{l_2(u)}\right)\circ \left(d\log{\left(-\frac{u_{kl,T}^2}{u_{kl}^2}\right)}\circ\right)^{i-1}d\log{\left(\frac{u_{kl,-}u_{kl,+}}{u_{kl}^2}\right)}}\,.\label{eq:Ikliiterinteg}
\end{aligned}
\end{equation}
Here, we define the \textit{alphabet} $\mathbb{A}$, which consists of the following six symbol \textit{letters}:
\begin{equation}
\begin{aligned}
\mathbb{A}=&\left\{l_1(u),\ldots,l_6(u)\right\}\\
=&\Bigg\{u-\frac{1}{2}+\frac{1}{2}\sqrt{\frac{v_{kl}^{(+)}}{v_{kl}^{(-)}}},\;-u+\frac{1}{2}+\frac{1}{2}\sqrt{\frac{v_{kl}^{(+)}}{v_{kl}^{(-)}}},\\
&u+\frac{\harpoon{v}_{k,T}\cdot \harpoon{v}_{l,T}-v_{l,T}^2+\sqrt{\left(\harpoon{v}_{k,T}\cdot \harpoon{v}_{l,T}\right)^2-v_{k,T}^2v_{l,T}^2}}{v_{k,T}^2+v_{l,T}^2-2\harpoon{v}_{k,T}\cdot \harpoon{v}_{l,T}},\\
&u+\frac{\harpoon{v}_{k,T}\cdot \harpoon{v}_{l,T}-v_{l,T}^2-\sqrt{\left(\harpoon{v}_{k,T}\cdot \harpoon{v}_{l,T}\right)^2-v_{k,T}^2v_{l,T}^2}}{v_{k,T}^2+v_{l,T}^2-2\harpoon{v}_{k,T}\cdot \harpoon{v}_{l,T}},\\
&u+\frac{v_{l,-}}{v_{k,-}-v_{l,-}},\;u+\frac{v_{l,+}}{v_{k,+}-v_{l,+}}\Bigg\}\,.
\end{aligned}
\end{equation}
Owing to the identities
\begin{eqnarray}
d\log{\left(-\frac{u_{kl,T}^2}{u_{kl}^2}\right)}&=&d\log{l_3(u)}+d\log{l_4(u)}-d\log{l_1(u)}-d\log{l_2(u)}\,,\\
d\log{\left(\frac{u_{kl,-}u_{kl,+}}{u_{kl}^2}\right)}&=&d\log{l_5(u)}+d\log{l_6(u)}-d\log{l_1(u)}-d\log{l_2(u)}\,,
\end{eqnarray}
the right-hand side of eq.~\eqref{eq:Ikliiterinteg} expands into $2^{2i+1}$ iterated integrals. For these iterated path integrals, we still need to impose the boundary conditions, which we choose at $u=0$. In this case, only the following four one-fold iterated integrals are relevant:
\begin{eqnarray}
\left.\int{d\log{l_5(u)}}\right|_{u=0}&=&\left.\log{u_{kl,-}}\right|_{u=0}=\log{\left(v_{l,-}\right)}\,,\\
\left.\int{d\log{l_6(u)}}\right|_{u=0}&=&\left.\log{u_{kl,+}}\right|_{u=0}=\log{\left(v_{l,+}\right)}\,,\\
\left.\int{d\log{l_1(u)}}\right|_{u=0}&=&\left.\int{d\log{l_2(u)}}\right|_{u=0}=0\,.
\end{eqnarray}
Additionally, for the $j$-fold iterated integrals, we have
\begin{equation}
\left.\int{\left(d\log{l_3(u)}\circ\right)^{j-1}d\log{l_5(u)}}\right|_{u=0}=-\mathrm{Li}_{j}(-v_{l,T}^2),\quad \mathrm{for}~2\leq j\leq i-1\,,\label{eq:BCl3l5}
\end{equation}
which follows from the integration in equation \eqref{eq:Ikliiterinteg} at the boundary point $u=0$. This boundary condition is well-defined because the letter $l_3(u)$ ($l_5(u)$) always appears together with $l_4(u)$ ($l_6(u)$). The iterated path integrals with these boundary conditions can be systematically reexpressed in terms of GPLs.
The expression for $I_{kl}^{(1)}$ in terms of GPLs is given by
\begin{align}
I_{kl}^{(1)}=&\Bigg[-G\left(-l_1,-l_1;1\right)-G\left(-l_1,l_2;1\right)+G\left(-l_1,-l_5;1\right)+G\left(-l_1,-l_6;1\right)\nonumber\\
&+G\left(l_2,-l_1;1\right)+G\left(l_2,l_2;1\right)-G\left(l_2,-l_5;1\right)-G\left(l_2,-l_6;1\right)\nonumber\\
&+\left(G\left(-l_1;1\right)-G\left(l_2;1\right)\right) \log{\left(v_{l,+} v_{l,-}\right)}\Bigg]\times\frac{1}{2\sqrt{v_{kl}^{(+)}v_{kl}^{(-)}}}\,,
\end{align}
where we use the shorthand notation $l_k=l_k(u=0)$. The transcendentality weight $3$ counterpart is given by:
\begin{equation}
\begin{aligned}
I_{kl}^{(2)}=&\Bigg[\Bigg(G\left(-l_1,-l_1,-l_1;1\right)+G\left(-l_1,-l_1,l_2;1\right)-G\left(-l_1,-l_1,-l_5;1\right)-G\left(-l_1,-l_1,-l_6;1\right)\\
&+G\left(-l_1,l_2,-l_1;1\right)+G\left(-l_1,l_2,l_2;1\right)-G\left(-l_1,l_2,-l_5;1\right)-G\left(-l_1,l_2,-l_6;1\right)\\
&-G\left(-l_1,-l_3,-l_1;1\right)-G\left(-l_1,-l_3,l_2;1\right)+G\left(-l_1,-l_3,-l_5;1\right)+G\left(-l_1,-l_3,-l_6;1\right)\\
&-G\left(-l_1,-l_4,-l_1;1\right)-G\left(-l_1,-l_4,l_2;1\right)+G\left(-l_1,-l_4,-l_5;1\right)+G\left(-l_1,-l_4,-l_6;1\right)\\
&-G\left(l_2,-l_1,-l_1;1\right)-G\left(l_2,-l_1,l_2;1\right)+G\left(l_2,-l_1,-l_5;1\right)+G\left(l_2,-l_1,-l_6;1\right)\\
&-G\left(l_2,l_2,-l_1;1\right)-G\left(l_2,l_2,l_2;1\right)+G\left(l_2,l_2,-l_5;1\right)+G\left(l_2,l_2,-l_6;1\right)\\
&+G\left(l_2,-l_3,-l_1;1\right)+G\left(l_2,-l_3,l_2;1\right)-G\left(l_2,-l_3,-l_5;1\right)-G\left(l_2,-l_3,-l_6;1\right)\\
&+G\left(l_2,-l_4,-l_1;1\right)+G\left(l_2,-l_4,l_2;1\right)-G\left(l_2,-l_4,-l_5;1\right)-G\left(l_2,-l_4,-l_6;1\right)\Bigg)\\
&+\Bigg(-G\left(-l_1,-l_1;1\right)-G\left(-l_1,l_2;1\right)+G\left(-l_1,-l_3;1\right)+G\left(-l_1,-l_4;1\right)\\
&+G\left(l_2,-l_1;1\right)+G\left(l_2,l_2;1\right)-G\left(l_2,-l_3;1\right)-G\left(l_2,-l_4;1\right)\Bigg)\log{\left(v_{l,+}v_{l,-}\right)}\\
&+\Bigg(G\left(l_2;1\right)-G\left(-l_1;1\right)\Bigg)\mathrm{Li}_2\left(-v_{l,T}^2\right)\Bigg]\times\frac{1}{2\sqrt{v_{kl}^{(+)}v_{kl}^{(-)}}}\,.
\end{aligned}
\end{equation}
The weight $4$ term is given by:
\begin{align}
I_{kl}^{(3)}=&\Bigg[\Bigg(-G\left(-l_1,-l_1,-l_1,-l_1;1\right)-G\left(-l_1,-l_1,-l_1,l_2;1\right)+G\left(-l_1,-l_1,-l_1,-l_5;1\right)\nonumber\\
&+G\left(-l_1,-l_1,-l_1,-l_6;1\right)-G\left(-l_1,-l_1,l_2,-l_1;1\right)-G\left(-l_1,-l_1,l_2,l_2;1\right)\nonumber\\
&+G\left(-l_1,-l_1,l_2,-l_5;1\right)+G\left(-l_1,-l_1,l_2,-l_6;1\right)+G\left(-l_1,-l_1,-l_3,-l_1;1\right)\nonumber\\
&+G\left(-l_1,-l_1,-l_3,l_2;1\right)-G\left(-l_1,-l_1,-l_3,-l_5;1\right)-G\left(-l_1,-l_1,-l_3,-l_6;1\right)\nonumber\\
&+G\left(-l_1,-l_1,-l_4,-l_1;1\right)+G\left(-l_1,-l_1,-l_4,l_2;1\right)-G\left(-l_1,-l_1,-l_4,-l_5;1\right)\nonumber\\
&-G\left(-l_1,-l_1,-l_4,-l_6;1\right)-G\left(-l_1,l_2,-l_1,-l_1;1\right)-G\left(-l_1,l_2,-l_1,l_2;1\right)\nonumber\\
&+G\left(-l_1,l_2,-l_1,-l_5;1\right)+G\left(-l_1,l_2,-l_1,-l_6;1\right)-G\left(-l_1,l_2,l_2,-l_1;1\right)\nonumber\\
&-G\left(-l_1,l_2,l_2,l_2;1\right)+G\left(-l_1,l_2,l_2,-l_5;1\right)+G\left(-l_1,l_2,l_2,-l_6;1\right)\nonumber\\
&+G\left(-l_1,l_2,-l_3,-l_1;1\right)+G\left(-l_1,l_2,-l_3,l_2;1\right)-G\left(-l_1,l_2,-l_3,-l_5;1\right)\nonumber\\
&-G\left(-l_1,l_2,-l_3,-l_6;1\right)+G\left(-l_1,l_2,-l_4,-l_1;1\right)+G\left(-l_1,l_2,-l_4,l_2;1\right)\nonumber\\
&-G\left(-l_1,l_2,-l_4,-l_5;1\right)-G\left(-l_1,l_2,-l_4,-l_6;1\right)+G\left(-l_1,-l_3,-l_1,-l_1;1\right)\nonumber\\
&+G\left(-l_1,-l_3,-l_1,l_2;1\right)-G\left(-l_1,-l_3,-l_1,-l_5;1\right)-G\left(-l_1,-l_3,-l_1,-l_6;1\right)\nonumber\\
&+G\left(-l_1,-l_3,l_2,-l_1;1\right)+G\left(-l_1,-l_3,l_2,l_2;1\right)-G\left(-l_1,-l_3,l_2,-l_5;1\right)\nonumber\\
&-G\left(-l_1,-l_3,l_2,-l_6;1\right)-G\left(-l_1,-l_3,-l_3,-l_1;1\right)-G\left(-l_1,-l_3,-l_3,l_2;1\right)\nonumber\\
&+G\left(-l_1,-l_3,-l_3,-l_5;1\right)+G\left(-l_1,-l_3,-l_3,-l_6;1\right)-G\left(-l_1,-l_3,-l_4,-l_1;1\right)\nonumber\\
&-G\left(-l_1,-l_3,-l_4,l_2;1\right)+G\left(-l_1,-l_3,-l_4,-l_5;1\right)+G\left(-l_1,-l_3,-l_4,-l_6;1\right)\nonumber\\
&+G\left(-l_1,-l_4,-l_1,-l_1;1\right)+G\left(-l_1,-l_4,-l_1,l_2;1\right)-G\left(-l_1,-l_4,-l_1,-l_5;1\right)\nonumber\\
&-G\left(-l_1,-l_4,-l_1,-l_6;1\right)+G\left(-l_1,-l_4,l_2,-l_1;1\right)+G\left(-l_1,-l_4,l_2,l_2;1\right)\nonumber\\
&-G\left(-l_1,-l_4,l_2,-l_5;1\right)-G\left(-l_1,-l_4,l_2,-l_6;1\right)-G\left(-l_1,-l_4,-l_3,-l_1;1\right)\nonumber\\
&-G\left(-l_1,-l_4,-l_3,l_2;1\right)+G\left(-l_1,-l_4,-l_3,-l_5;1\right)+G\left(-l_1,-l_4,-l_3,-l_6;1\right)\nonumber\\
&-G\left(-l_1,-l_4,-l_4,-l_1;1\right)-G\left(-l_1,-l_4,-l_4,l_2;1\right)+G\left(-l_1,-l_4,-l_4,-l_5;1\right)\nonumber\\
&+G\left(-l_1,-l_4,-l_4,-l_6;1\right)+G\left(l_2,-l_1,-l_1,-l_1;1\right)+G\left(l_2,-l_1,-l_1,l_2;1\right)\nonumber\\
&-G\left(l_2,-l_1,-l_1,-l_5;1\right)-G\left(l_2,-l_1,-l_1,-l_6;1\right)+G\left(l_2,-l_1,l_2,-l_1;1\right)\nonumber\\
&+G\left(l_2,-l_1,l_2,l_2;1\right)-G\left(l_2,-l_1,l_2,-l_5;1\right)-G\left(l_2,-l_1,l_2,-l_6;1\right)\nonumber\\
&-G\left(l_2,-l_1,-l_3,-l_1;1\right)-G\left(l_2,-l_1,-l_3,l_2;1\right)+G\left(l_2,-l_1,-l_3,-l_5;1\right)\nonumber\\
&+G\left(l_2,-l_1,-l_3,-l_6;1\right)-G\left(l_2,-l_1,-l_4,-l_1;1\right)-G\left(l_2,-l_1,-l_4,l_2;1\right)\nonumber\\
&+G\left(l_2,-l_1,-l_4,-l_5;1\right)+G\left(l_2,-l_1,-l_4,-l_6;1\right)+G\left(l_2,l_2,-l_1,-l_1;1\right)\nonumber\\
&+G\left(l_2,l_2,-l_1,l_2;1\right)-G\left(l_2,l_2,-l_1,-l_5;1\right)-G\left(l_2,l_2,-l_1,-l_6;1\right)\nonumber\\
&+G\left(l_2,l_2,l_2,-l_1;1\right)+G\left(l_2,l_2,l_2,l_2;1\right)-G\left(l_2,l_2,l_2,-l_5;1\right)\nonumber\\
&-G\left(l_2,l_2,l_2,-l_6;1\right)-G\left(l_2,l_2,-l_3,-l_1;1\right)-G\left(l_2,l_2,-l_3,l_2;1\right)\nonumber\\
&+G\left(l_2,l_2,-l_3,-l_5;1\right)+G\left(l_2,l_2,-l_3,-l_6;1\right)-G\left(l_2,l_2,-l_4,-l_1;1\right)\nonumber\\
&-G\left(l_2,l_2,-l_4,l_2;1\right)+G\left(l_2,l_2,-l_4,-l_5;1\right)+G\left(l_2,l_2,-l_4,-l_6;1\right)\nonumber\\
&-G\left(l_2,-l_3,-l_1,-l_1;1\right)-G\left(l_2,-l_3,-l_1,l_2;1\right)+G\left(l_2,-l_3,-l_1,-l_5;1\right)\nonumber\\
&+G\left(l_2,-l_3,-l_1,-l_6;1\right)-G\left(l_2,-l_3,l_2,-l_1;1\right)-G\left(l_2,-l_3,l_2,l_2;1\right)\nonumber\\
&+G\left(l_2,-l_3,l_2,-l_5;1\right)+G\left(l_2,-l_3,l_2,-l_6;1\right)+G\left(l_2,-l_3,-l_3,-l_1;1\right)\nonumber\\
&+G\left(l_2,-l_3,-l_3,l_2;1\right)-G\left(l_2,-l_3,-l_3,-l_5;1\right)-G\left(l_2,-l_3,-l_3,-l_6;1\right)\nonumber\\
&+G\left(l_2,-l_3,-l_4,-l_1;1\right)+G\left(l_2,-l_3,-l_4,l_2;1\right)-G\left(l_2,-l_3,-l_4,-l_5;1\right)\nonumber\\
&-G\left(l_2,-l_3,-l_4,-l_6;1\right)-G\left(l_2,-l_4,-l_1,-l_1;1\right)-G\left(l_2,-l_4,-l_1,l_2;1\right)\nonumber\\
&+G\left(l_2,-l_4,-l_1,-l_5;1\right)+G\left(l_2,-l_4,-l_1,-l_6;1\right)-G\left(l_2,-l_4,l_2,-l_1;1\right)\nonumber\\
&-G\left(l_2,-l_4,l_2,l_2;1\right)+G\left(l_2,-l_4,l_2,-l_5;1\right)+G\left(l_2,-l_4,l_2,-l_6;1\right)\nonumber\\
&+G\left(l_2,-l_4,-l_3,-l_1;1\right)+G\left(l_2,-l_4,-l_3,l_2;1\right)-G\left(l_2,-l_4,-l_3,-l_5;1\right)\nonumber\\
&-G\left(l_2,-l_4,-l_3,-l_6;1\right)+G\left(l_2,-l_4,-l_4,-l_1;1\right)+G\left(l_2,-l_4,-l_4,l_2;1\right)\nonumber\\
&-G\left(l_2,-l_4,-l_4,-l_5;1\right)-G\left(l_2,-l_4,-l_4,-l_6;1\right)\Bigg)\nonumber\\
&+\Bigg(G\left(-l_1,-l_1,-l_1;1\right)+G\left(-l_1,-l_1,l_2;1\right)-G\left(-l_1,-l_1,-l_3;1\right)-G\left(-l_1,-l_1,-l_4;1\right)\nonumber\\
&+G\left(-l_1,l_2,-l_1;1\right)+G\left(-l_1,l_2,l_2;1\right)-G\left(-l_1,l_2,-l_3;1\right)-G\left(-l_1,l_2,-l_4;1\right)\nonumber\\
&-G\left(-l_1,-l_3,-l_1;1\right)-G\left(-l_1,-l_3,l_2;1\right)+G\left(-l_1,-l_3,-l_3;1\right)+G\left(-l_1,-l_3,-l_4;1\right)\nonumber\\
&-G\left(-l_1,-l_4,-l_1;1\right)-G\left(-l_1,-l_4,l_2;1\right)+G\left(-l_1,-l_4,-l_3;1\right)+G\left(-l_1,-l_4,-l_4;1\right)\nonumber\\
&-G\left(l_2,-l_1,-l_1;1\right)-G\left(l_2,-l_1,l_2;1\right)+G\left(l_2,-l_1,-l_3;1\right)+G\left(l_2,-l_1,-l_4;1\right)\nonumber\\
&-G\left(l_2,l_2,-l_1;1\right)-G\left(l_2,l_2,l_2;1\right)+G\left(l_2,l_2,-l_3;1\right)+G\left(l_2,l_2,-l_4;1\right)\nonumber\\
&+G\left(l_2,-l_3,-l_1;1\right)+G\left(l_2,-l_3,l_2;1\right)-G\left(l_2,-l_3,-l_3;1\right)-G\left(l_2,-l_3,-l_4;1\right)\nonumber\\
&+G\left(l_2,-l_4,-l_1;1\right)+G\left(l_2,-l_4,l_2;1\right)-G\left(l_2,-l_4,-l_3;1\right)-G\left(l_2,-l_4,-l_4;1\right)\Bigg) \log{\left(v_{l,+} v_{l,-}\right)}\nonumber\\
&+\Bigg(G\left(-l_1,-l_1;1\right)+G\left(-l_1,l_2;1\right)-G\left(-l_1,-l_3;1\right)-G\left(-l_1,-l_4;1\right)\nonumber\\
&-G\left(l_2,-l_1;1\right)-G\left(l_2,l_2;1\right)+G\left(l_2,-l_3;1\right)+G\left(l_2,-l_4;1\right)\Bigg)
   \mathrm{Li}_2\left(-v_{l,T}^2\right)\nonumber\\
&+\Bigg(G\left(l_2;1\right)-G\left(-l_1;1\right)\Bigg)\mathrm{Li}_3\left(-v_{l,T}^2\right)\Bigg]\times \frac{1}{2\sqrt{v_{kl}^{(+)}v_{kl}^{(-)}}}\,.
\end{align}
We can systematically generate $I_{kl}^{(i)}$ for $i\geq 4$ in a similar manner. However, their expressions are too lengthy to be presented here. Specifically, the numbers of GPLs in $I_{kl}^{(1)},I_{kl}^{(2)},I_{kl}^{(3)}$, and $I_{kl}^{(4)}$ are $10,42,170$, and $682$, respectively. The explicit expressions for $I_{kl}^{(i)}$ with $1\leq i\leq 4$ are available in the {\sc\small Mathematica} ancillary file\\
{\tt TwoMassiveEikonal\_AzimuthAveraged.wl}, which is submitted along with this paper. As in the previous cases, the azimuthally averaged TMD eikonal integral has no imaginary component.

We compare our analytic expressions for $I_{kl}^{(i)}$ with direct numerical integration over $u$ using the {\sc\small Mathematica} function {\tt NIntegrate} for more than 2000 randomly generated phase-space points, and we find perfect agreement. However, for the numerical evaluation of GPLs in $I_{kl}^{(i)}$, we need to use both {\tt handyG} and  {\tt PolyLogTools}, as each has different limitations. Some GPL values at specific phase-space points turn out to be incorrect when using a single tool. We have also checked our one-loop azimuthally-averaged soft function against eqs.(3.5-3.7) in ref.~\cite{Angeles-Martinez:2018mqh} for back-to-back heavy quark pair production up to $\mathcal{O}(\epsilon)$.

\subsubsection{Azimuthal-angle-dependent result}

The TMD eikonal integral with the full azimuthal-angle dependence is given by
\begin{eqnarray}
\bar{\mathcal{E}}_{\perp,kl}^{(m_k,m_l)}(\harpoon{b}_{T},\mu)&=&8\pi^2\mu^{2\epsilon}\int{\frac{dk_{g,-}d^{d-2}\harpoon{k}_{g,T}}{\left(2\pi\right)^{d-1}}\frac{1}{2k_{g,-}}\frac{v_k\cdot v_l}{\left(v_k\cdot k_g\right)\left(v_l\cdot k_g\right)}e^{-i\harpoon{k}_{g,T}\cdot \harpoon{b}_{T}}}\nonumber\\
&=&8\pi^2\mu^{2\epsilon}\int_0^1{du \int{\frac{dk_{g,-}d^{d-2}\harpoon{k}_{g,T}}{\left(2\pi\right)^{d-1}}\frac{1}{2k_{g,-}}\frac{v_k\cdot v_l}{\left[u v_k\cdot k_g+(1-u)v_l\cdot k_g\right]^2}e^{-i\harpoon{k}_{g,T}\cdot \harpoon{b}_{T}}}}\nonumber\\
&=&\pi^\epsilon \left(b_T\mu\right)^{2\epsilon} \int_0^{1}{du\frac{v_k\cdot v_l}{u_{kl}^2}\left[\Gamma\left(-\epsilon\right){}_2\hspace{-0.7mm}F_1\left(1,-\epsilon;\frac{1}{2};-\frac{\left(\harpoon{u}_{kl,T}\cdot \harpoon{b}_{T}\right)^2}{b_T^2u_{kl}^2}\right)\right.}\nonumber\\
&&\left.-i\sqrt{\pi}\Gamma\left(\frac{1}{2}-\epsilon\right)\frac{\left(\harpoon{u}_{kl,T}\cdot \harpoon{b}_{T}\right)}{b_Tu_{kl}}\left(\frac{\left(\harpoon{u}_{kl,T}\cdot \harpoon{b}_{T}\right)^2}{b_T^2 u_{kl}^2}+1\right)^{-\frac{1}{2}+\epsilon}\right]\,,\label{eq:twomassiveoneloopbTsoft}
\end{eqnarray}
where, in the second equation, we have used the Feynman parameterization from eq.~\eqref{eq:Feynmanparameter}, and introduced the shorthand notation $u_{kl}^\mu \equiv uv_k^\mu+(1-u)v_l^\mu$ and its transverse vector $\harpoon{u}_{kl,T}$, such that $u_{kl}=\sqrt{u_{kl}\cdot u_{kl}}=\sqrt{u^2+(1-u)^2+2u(1-u)v_k\cdot v_l}$.

Next, we calculate the analytical expression of eq.~\eqref{eq:twomassiveoneloopbTsoft} in terms of the Laurent series expansion in the dimensional regulator $\epsilon$. With the help of the {\sc\small Mathematica} package  {\tt HypExp}~\cite{Huber:2007dx}, we can reexpress the two-massive TMD eikonal integral in eq.~\eqref{eq:twomassiveoneloopbTsoft} as
\begin{eqnarray}
\bar{\mathcal{E}}_{\perp,kl}^{(m_k,m_l)}(\harpoon{b}_{T})&=&\frac{(4\pi)^\epsilon}{\Gamma(1-\epsilon)}v_k\cdot v_l\Bigg[-\frac{I^{(0)}_{kl}}{\epsilon}-I^{(0)}_{kl} \left(\sum_{j=1}^{+\infty}{\epsilon^{j-1}\frac{L^j_\perp}{j!}}\right)+ \sum_{j=0}^{+\infty}{\epsilon^j J_j}\Bigg]\,,
\end{eqnarray}
where $I^{(0)}_{kl}$ is given in eq.~\eqref{eq:Ikl0exp1}. All $J_j$'s are one-fold integrals yet to be determined. The first four lowest-order terms in $\epsilon$ are
\begin{eqnarray}
J_0&=&\int_0^1{du \frac{x_{u,b}}{u_{kl}^2}\left[\log{\left(\frac{1-x_{u,b}}{1+x_{u,b}}\right)}-i\pi \theta_{u,b}\right]}\equiv \tilde{J}_0\,,\label{eq:J0integralsdef1}\\
J_1&=&L_\perp J_0-\zeta_2I_{kl}^{(0)}+\int_0^1{du \frac{x_{u,b}}{u_{kl}^2}\Bigg[\left(\mathrm{Li}_2\left(\frac{1+x_{u,b}}{2}\right)-\mathrm{Li}_2\left(\frac{1-x_{u,b}}{2}\right)\right)}\nonumber\\
&&+\frac{1}{2}\left(\log^2{\left(\frac{1+x_{u,b}}{2}\right)}-\log^2{\left(\frac{1-x_{u,b}}{2}\right)}\right)+i\pi \theta_{u,b}\log{\left(\frac{1-x_{u,b}^2}{4}\right)}\Bigg]\nonumber\\
&\equiv&L_\perp J_0-\zeta_2I_{kl}^{(0)}+\tilde{J}_1\,,\label{eq:J1integralsdef1}\\
J_2&=&\frac{L_\perp^2}{2!} J_0 + L_\perp \left(\left.J_1\right|_{L_\perp \to 0}\right)-\frac{2}{3}\zeta_3I_{kl}^{(0)}+\int_0^{1}{du \frac{x_{u,b}}{u_{kl}^2}\Bigg[2\left(\mathrm{Li}_3\left(\frac{1-x_{u,b}}{2}\right)-\mathrm{Li}_3\left(\frac{1+x_{u,b}}{2}\right)\right)}\nonumber\\
&&+\frac{1}{6}\left(\log^3{\left(\frac{1-x_{u,b}}{2}\right)}-\log^3{\left(\frac{1+x_{u,b}}{2}\right)}\right)+\frac{1}{2}\log{\left(\frac{1+x_{u,b}}{2}\right)}\log^2{\left(\frac{1-x_{u,b}}{2}\right)}\nonumber\\
&&-\frac{1}{2}\log^2{\left(\frac{1+x_{u,b}}{2}\right)}\log{\left(\frac{1-x_{u,b}}{2}\right)}-i\pi\theta_{u,b}\bigg(\frac{1}{2}\log^2{\left(\frac{1-x_{u,b}^2}{4}\right)}+2\zeta_2\bigg)\Bigg]\nonumber\\
&\equiv&\frac{L_\perp^2}{2!} J_0 + L_\perp \left(\left.J_1\right|_{L_\perp \to 0}\right)-\frac{2}{3}\zeta_3I_{kl}^{(0)}+\tilde{J}_2\,,\label{eq:J2integralsdef1}\\
J_3&=& \frac{L_\perp^3}{3!} J_0 + \frac{L_\perp^2}{2!} \left(\left.J_1\right|_{L_\perp \to 0}\right)+L_\perp \left(\left.J_2\right|_{L_\perp \to 0}\right)-\frac{7}{4}\zeta_4I_{kl}^{(0)}\nonumber\\
&&+\int_0^{1}{du \frac{x_{u,b}}{u_{kl}^2}\Bigg[2\left(\mathrm{Li}_4\left(\frac{1+x_{u,b}}{2}\right)-\mathrm{Li}_4\left(\frac{1-x_{u,b}}{2}\right)\right)+\left(\mathrm{S}_{2,2}\left(\frac{1+x_{u,b}}{2}\right)-\mathrm{S}_{2,2}\left(\frac{1-x_{u,b}}{2}\right)\right)}\nonumber\\
&&+\left(\log{\left(\frac{1-x_{u,b}}{2}\right)}\mathrm{Li}_3\left(\frac{1+x_{u,b}}{2}\right)-\log{\left(\frac{1+x_{u,b}}{2}\right)}\mathrm{Li}_3\left(\frac{1-x_{u,b}}{2}\right)\right)\nonumber\\
&&+\zeta_2\left(\mathrm{Li}_2\left(\frac{1+x_{u,b}}{2}\right)-\mathrm{Li}_2\left(\frac{1-x_{u,b}}{2}\right)\right)-\frac{5}{3}\zeta_3\log{\left(\frac{1+x_{u,b}}{1-x_{u,b}}\right)}\nonumber\\
&&+\frac{1}{24}\log{\left(\frac{1+x_{u,b}}{1-x_{u,b}}\right)}\log{\left(\frac{1-x_{u,b}^2}{4}\right)}\left(\log^2{\left(\frac{1-x_{u,b}^2}{4}\right)}+2\log{\left(\frac{1+x_{u,b}}{2}\right)}\log{\left(\frac{1-x_{u,b}}{2}\right)}\right)\nonumber\\
&&+i\pi\theta_{u,b}\bigg(\frac{1}{6}\log^3{\left(\frac{1-x_{u,b}^2}{4}\right)}+2\zeta_2\log{\left(\frac{1-x_{u,b}^2}{4}\right)}-\frac{8}{3}\zeta_3\bigg)\Bigg]\nonumber\\
&\equiv&\frac{L_\perp^3}{3!} J_0 + \frac{L_\perp^2}{2!} \left(\left.J_1\right|_{L_\perp \to 0}\right)+L_\perp \left(\left.J_2\right|_{L_\perp \to 0}\right)-\frac{7}{4}\zeta_4I_{kl}^{(0)}+\tilde{J}_3\,,\label{eq:J3integralsdef1}
\end{eqnarray}
where
\begin{eqnarray}
x_{u,b}&\equiv& \sqrt{\frac{\left(\harpoon{u}_{kl,T}\cdot \hat{\harpoon{b}}_T\right)^2}{\left(\harpoon{u}_{kl,T}\cdot \hat{\harpoon{b}}_T\right)^2+u_{kl}^2}}\,,\\
\theta_{u,b}&\equiv&\frac{\harpoon{u}_{kl,T}\cdot \hat{\harpoon{b}}_T}{|\harpoon{u}_{kl,T}\cdot \hat{\harpoon{b}}_T|}\,
\end{eqnarray}
with the unit $(d-2)$-dimensional vector $\hat{\harpoon{b}}_T\equiv \harpoon{b}_T/b_T$.
In principle, the integrals $J_i$ can be evaluated using numerical integration methods. Thus, ref.~\cite{Catani:2021cbl} (cf. eqs.(25,26) therein) leaves these one-fold integrals untouched. Here, however, we aim to solve them in terms of multiple polylogarithms. 

We notice that the integrands of all $\tilde{J}_i$'s are invariant under the following replacement:
\begin{equation}
x_{u,b}\to -x_{u,b},\quad \theta_{u,b}\to -\theta_{u,b}\,.
\end{equation}
The main complication arises from the square root in $x_{u,b}$, which we must rationalize. As a first step, we can rewrite:
\begin{equation}
x_{u,b}=\frac{1}{\theta_{u,b}}\frac{\harpoon{u}_{kl,T}\cdot \hat{\harpoon{b}}_T}{\sqrt{\left(\harpoon{u}_{kl,T}\cdot \hat{\harpoon{b}}_T\right)^2+u_{kl}^2}}\,.
\end{equation}
In this case, the integrand of $\tilde{J}_i$ is symmetric for the replacement $\theta_{u,b}\to -\theta_{u,b}$. Therefore, we can assume $\theta_{u,b}=1$, as we can always change $\theta_{u,b}\to -\theta_{u,b}$. The only square root that needs to be rationalized is:
\begin{equation}
\sqrt{\left(\harpoon{u}_{kl,T}\cdot \hat{\harpoon{b}}_T\right)^2+u_{kl}^2}=\sqrt{\left(\underbrace{\left(\harpoon{v}_{k,T}\cdot \hat{\harpoon{b}}_T-\harpoon{v}_{l,T}\cdot \hat{\harpoon{b}}_T\right)^2+2-2v_k\cdot v_l}_{\equiv -A_{kl,b}}\right)\left(u-u_+\right)\left(u-u_-\right)}\,,
\end{equation}
where the roots are
\begin{equation}
\begin{aligned}
u_\pm=&\frac{v_k\cdot v_l-1+\left(\harpoon{v}_{k,T}\cdot \hat{\harpoon{b}}_T-\harpoon{v}_{l,T}\cdot \hat{\harpoon{b}}_T\right)\harpoon{v}_{l,T}\cdot \hat{\harpoon{b}}_T}{A_{kl,b}}\\
&\pm \frac{\sqrt{(v_k\cdot v_l-1)(1+v_k\cdot v_l+2\harpoon{v}_{k,T}\cdot \hat{\harpoon{b}}_T \harpoon{v}_{l,T}\cdot \hat{\harpoon{b}}_T)-\left(\harpoon{v}_{k,T}\cdot \hat{\harpoon{b}}_T-\harpoon{v}_{l,T}\cdot \hat{\harpoon{b}}_T\right)^2}}{A_{kl,b}}\,.
\end{aligned}
\end{equation}
We can generally assume that $u_{\pm}\in\mathbb{R}$. In order to rationalize $\sqrt{-A_{kl,b}(u-u_+)(u-u_-)}$, let us first use the substitution $u=(u_+-u_-)x+u_+$ to obtain:
\begin{equation}
\sqrt{-A_{kl,b}(u-u_+)(u-u_-)}=\sqrt{-A_{kl,b}(u_+-u_-)^2x(x+1)}\,.
\end{equation}
The last square root can be further rationalized using the package {\tt RationalizeRoots}~\cite{Besier:2019kco}. By making the change of variable $x=-1/(y^2+1)$, the square root becomes:
\begin{equation}
\sqrt{-A_{kl,b}(u_+-u_-)^2x(x+1)}=\sqrt{A_{kl,b}(u_+-u_-)^2\left(\frac{y}{1+y^2}\right)^2}=\sqrt{A_{kl,b}}|u_+-u_-|\frac{u_+}{|u_+|}\frac{y}{1+y^2}\,.
\end{equation}
In fact, we have:
\begin{equation}
y=\sqrt{\frac{u-u_-}{u_+-u}}\,,\quad u=\frac{u_--u_+}{y^2+1}+u_+\,.
\end{equation}
Additionally, we can express:
\begin{equation}
\begin{aligned}
x_{u,b}=&\frac{1}{\sqrt{A_{kl,b}}}\Bigg[\underbrace{\frac{u_+}{|u_+|}\frac{u_-\harpoon{v}_{k,T}\cdot \hat{\harpoon{b}}_T+(1-u_-)\harpoon{v}_{l,T}\cdot \hat{\harpoon{b}}_T}{|u_+-u_-|}}_{\equiv x_{kl,-}}\frac{1}{y}\\
&+\underbrace{\frac{u_+}{|u_+|}\frac{u_+\harpoon{v}_{k,T}\cdot \hat{\harpoon{b}}_T+(1-u_+)\harpoon{v}_{l,T}\cdot \hat{\harpoon{b}}_T}{|u_+-u_-|}}_{\equiv x_{kl,+}}y\Bigg]\,.
\end{aligned}
\end{equation}
Therefore, the partial fractions give us:
\begin{equation}
\begin{aligned}
\frac{du}{u_{kl}^2}=&\frac{du}{2(1-v_k\cdot v_l)}\frac{1}{\left(u-\frac{1}{2}+\frac{1}{2}\sqrt{\frac{v_{kl}^{(+)}}{v_{kl}^{(-)}}}\right)\left(u-\frac{1}{2}-\frac{1}{2}\sqrt{\frac{v_{kl}^{(+)}}{v_{kl}^{(-)}}}\right)}\\
=&\frac{du}{2\sqrt{v_{kl}^{(+)}v_{kl}^{(-)}}}\left[\frac{1}{u-u_{kl,-}}+\frac{1}{u_{kl,+}-u}\right]\\
=&\frac{dy}{\sqrt{v_{kl}^{(+)}v_{kl}^{(-)}}}\left(u_+-u_-\right)y\\
&\times\sum_{r=\pm}{\Bigg[\frac{r}{u_{kl,r}-u_+}\frac{1}{2(a_r+1)}\bigg(-\frac{i}{y+i}+\frac{i}{y-i}+\frac{1}{\sqrt{a_r}}\frac{1}{y-\sqrt{a_r}}-\frac{1}{\sqrt{a_r}}\frac{1}{y+\sqrt{a_r}}\bigg)\Bigg]}\,,
\end{aligned}
\end{equation}
where we have defined:
\begin{eqnarray}
u_{kl,\pm}&\equiv& \frac{1}{2}\pm\frac{1}{2}\sqrt{\frac{v_{kl}^{(+)}}{v_{kl}^{(-)}}}\,,\nonumber\\
a_{\pm}&\equiv&\frac{u_{kl,\pm}-u_-}{u_+-u_{kl,\pm}}\,.
\end{eqnarray}
Similarly, we obtain:
\begin{equation}
\begin{aligned}
du\frac{x_{u,b}}{u_{kl}^2}=&dy\frac{u_+-u_-}{\sqrt{A_{kl,b}}\sqrt{v_{kl}^{(+)}v_{kl}^{(-)}}}\left(x_{kl,+}y^2+x_{kl,-}\right)\\
&\times\sum_{r=\pm}{\Bigg[\frac{r}{u_{kl,r}-u_+}\frac{1}{2(a_r+1)}\bigg(-\frac{i}{y+i}+\frac{i}{y-i}+\frac{1}{\sqrt{a_r}}\frac{1}{y-\sqrt{a_r}}-\frac{1}{\sqrt{a_r}}\frac{1}{y+\sqrt{a_r}}\bigg)\Bigg]}\\
=&dy\frac{u_+-u_-}{\sqrt{A_{kl,b}}\sqrt{v_{kl}^{(+)}v_{kl}^{(-)}}}\sum_{r=\pm}{\Bigg[\frac{r}{u_{kl,r}-u_+}\frac{1}{2(a_r+1)}\frac{x_{kl,-}+a_rx_{kl,+}}{\sqrt{a_r}}\left(\frac{1}{y-\sqrt{a_r}}-\frac{1}{y+\sqrt{a_r}}\right)\Bigg]}\,.
\end{aligned}
\end{equation}

Let us define the \textit{alphabet} $\mathbb{A}_b$ consisting of the following 11 symbols, or \textit{letters}:
\begin{eqnarray}
\mathbb{A}_b&=&\left\{b_1(y),\ldots,b_{11}(y)\right\}\nonumber\\
&=&\Bigg\{y,\;y+\sqrt{a_+},\;y-\sqrt{a_+},\;y+\sqrt{a_-},\;y-\sqrt{a_-},\nonumber\\
&&y+\sqrt{-\frac{x_{kl,-}}{x_{kl,+}}},\;y-\sqrt{-\frac{x_{kl,-}}{x_{kl,+}}},y+\frac{\sqrt{A_{kl,b}}+\sqrt{A_{kl,b}-4x_{kl,+}x_{kl,-}}}{2x_{kl,+}},\nonumber\\
&&y-\frac{\sqrt{A_{kl,b}}+\sqrt{A_{kl,b}-4x_{kl,+}x_{kl,-}}}{2x_{kl,+}},\;y+\frac{-\sqrt{A_{kl,b}}+\sqrt{A_{kl,b}-4x_{kl,+}x_{kl,-}}}{2x_{kl,+}},\nonumber\\
&&y-\frac{-\sqrt{A_{kl,b}}+\sqrt{A_{kl,b}-4x_{kl,+}x_{kl,-}}}{2x_{kl,+}}\Bigg\}\,.
\end{eqnarray}
The following terms appearing in $\tilde{J}_i$ can be written using the above letters:
\begin{eqnarray}
du\frac{x_{u,b}}{u_{kl}^2}&=&\frac{u_+-u_-}{\sqrt{A_{kl,b}}\sqrt{v_{kl}^{(+)}v_{kl}^{(-)}}}\Bigg[\frac{1}{u_{kl,+}-u_+}\frac{1}{2(a_++1)}\frac{x_{kl,-}+a_+x_{kl,+}}{\sqrt{a_+}}\left(d\log{b_3(y)}-d\log{b_2(y)}\right)\nonumber\\
&&-\frac{1}{u_{kl,-}-u_+}\frac{1}{2(a_-+1)}\frac{x_{kl,-}+a_-x_{kl,+}}{\sqrt{a_-}}\left(d\log{b_5(y)}-d\log{b_4(y)}\right)\Bigg]\,,\nonumber\\
\frac{1+x_{u,b}}{2}&=&\frac{x_{kl,+}}{2\sqrt{A_{kl,b}}}\frac{b_{8}(y)b_{11}(y)}{b_1(y)}\,,\quad d\log{\left(\frac{1+x_{u,b}}{2}\right)}=d\log{b_{8}(y)}+d\log{b_{11}(y)}-d\log{b_{1}(y)}\,,\nonumber\\
\frac{1-x_{u,b}}{2}&=&-\frac{x_{kl,+}}{2\sqrt{A_{kl,b}}}\frac{b_{9}(y)b_{10}(y)}{b_1(y)}\,,\quad d\log{\left(\frac{1-x_{u,b}}{2}\right)}=d\log{b_{9}(y)}+d\log{b_{10}(y)}-d\log{b_{1}(y)}\,,\nonumber\\
\frac{x_{u,b}}{2}&=&\frac{x_{kl,+}}{2\sqrt{A_{kl,b}}}\frac{b_6(y)b_7(y)}{b_1(y)}\,,\quad d\log{\left(\frac{x_{u,b}}{2}\right)}=d\log{b_{6}(y)}+d\log{b_{7}(y)}-d\log{b_{1}(y)}\,.
\end{eqnarray}

We still need to fix the boundary conditions at the boundary point 
\begin{equation}
y=y_0\equiv \left.\sqrt{\frac{u-u_-}{u_+-u}}\right|_{u=0}=\sqrt{-\frac{u_-}{u_+}}\,.
\end{equation}
The boundary conditions can be specified as:
\begin{eqnarray}
\left.\int{d\log{b_i(y)}}\right|_{y=y_0}&=&0\,,\quad {\rm for}~i=1,2,3,4,5,7,10,11\,,\nonumber\\
\left.\int{d\log{b_{8}(y)}}\right|_{y=y_0}&=&\log{\left(\frac{1}{2}+\frac{x_{0,b}}{2}\right)}\,,\nonumber\\
\left.\int{d\log{b_{9}(y)}}\right|_{y=y_0}&=&\log{\left(\frac{1}{2}-\frac{x_{0,b}}{2}\right)}\,,\nonumber\\
\left.\int{d\log{b_{6}(y)}}\right|_{y=y_0}&=&\log{\left(\frac{x_{0,b}}{2}\right)}\,,\nonumber\\
\left.\int{\left(d\log{b_8(y)}\circ\right)^{j-1}d\log{b_9(y)}}\right|_{y=y_0}&=&-\mathrm{Li}_{j}\left(\frac{1}{2}+\frac{x_{0,b}}{2}\right)\,,\\
\left.\int{\left(d\log{b_9(y)}\circ\right)^{j-1}d\log{b_8(y)}}\right|_{y=y_0}&=&-\mathrm{Li}_{j}\left(\frac{1}{2}-\frac{x_{0,b}}{2}\right),\quad \mathrm{for}~j\geq 2\,,\nonumber\\
\left.\int{\prod_{r=1}^{r_{\mathrm{max}}}{\left[\left(d\log{b_8(y)}\circ\right)^{i_r}\left(d\log{b_9(y)}\circ\right)^{j_r}\right]}}\right|_{y=y_0}&=&G\left(\underbrace{0,\ldots,0}_{i_1},\underbrace{1,\ldots,1}_{j_1},\ldots,\underbrace{0,\ldots,0}_{i_{r_{\mathrm{max}}}},\underbrace{1,\ldots,1}_{j_{r_{\mathrm{max}}}};\frac{1+x_{0,b}}{2}\right)\nonumber\\
&=&G\left(\underbrace{1,\ldots,1}_{i_1},\underbrace{0,\ldots,0}_{j_1},\ldots,\underbrace{1,\ldots,1}_{i_{r_{\mathrm{max}}}},\underbrace{0,\ldots,0}_{j_{r_{\mathrm{max}}}};\frac{1-x_{0,b}}{2}\right)\,,\nonumber
\end{eqnarray}
where we have defined
\begin{equation}
x_{0,b}\equiv \left.x_{u,b}\right|_{u=0}=\frac{1}{\sqrt{A_{kl,b}}}\left(x_{kl,-}y_0^{-1}+x_{kl,+}y_0\right)\,.
\end{equation}
Inspecting the structure of $\tilde{J}_i$ in eqs.~\eqref{eq:J0integralsdef1}, \eqref{eq:J1integralsdef1}, \eqref{eq:J2integralsdef1}, and \eqref{eq:J3integralsdef1}, we actually only need to determine the boundary conditions for $d\log{\left(\frac{1\pm x_{u,b}}{2}\right)}$ at $y=y_0$. The boundary conditions for other iterated integrals will not be used. 

In order to express logarithms and multiple polylogarithms in terms of the iterated path integrals, we must ensure that the indices in $G()$ are independent of the variable $y$. Additionally, some extra low-weight integrals need to be added. The path of the iterated integrals starts at $y=y_0$ and ends at the point of $y=y_1+y_0$, with
\begin{equation}
y_1\equiv \sqrt{\frac{1-u_-}{u_+-1}}-y_0\,.
\end{equation}
When evaluating $G()$ numerically, we find that the last four letters $b_{8,9,10,11}(y)$ can always be mapped to the other four letters $b_{2,3,4,5}(y)$. However, their correspondence depends on the phase space or the values of the three independent variables $v_k\cdot v_l$, $\harpoon{v}_{k,T}\cdot \hat{\harpoon{b}}_T$, and $\harpoon{v}_{l,T}\cdot \hat{\harpoon{b}}_T$. To avoid numerical artifacts caused by round-off errors in floating-point calculations, we need to replace all $b_{8,9,10,11}$ with $b_{2,3,4,5}$ for each considered phase-space point. 

The final expressions for $\tilde{J}_i$ can be written in terms of GPLs with the indices $b_i\equiv b_i(y=y_0)$ and the argument $y_1$. The explicit analytic expression for $\tilde{J}_0$ is:
\begin{align}
\tilde{J}_0=&\Bigg\{\frac{x_{kl,-}+a_+ x_{kl,+}}{2\sqrt{a_+} \left(a_++1\right)\left(u_{kl,+}-u_+\right)}
   \Bigg[G\left(-b_2,-b_8;y_1\right)-G\left(-b_2,-b_9;y_1\right)-G\left(-b_2,-b_{10};y_1\right)\nonumber\\
   &+G\left(-b_2,-b_{11};y_1\right)-G\left(-b_3,-b_8;y_1\right)+G\left(-b_3,-b_9;y_1\right)+G\left(-b_3,-b_{10};y_1\right)\nonumber\\
   &-G\left(-b_3,-b_{11};y_1\right)-\log \left(\frac{1-x_{0,b}}{1+x_{0,b}}\right)
   \left(G\left(-b_2;y_1\right)-G\left(-b_3;y_1\right)\right)\Bigg]\nonumber\\
   &-\frac{x_{kl,-}+a_-x_{kl,+}}{2\sqrt{a_-} \left(a_-+1\right) \left(u_{kl,-}-u_+\right)}
   \Bigg[G\left(-b_4,-b_8;y_1\right)-G\left(-b_4,-b_9;y_1\right)-G\left(-b_4,-b_{10};y_1\right)\nonumber\\
   &+G\left(-b_4,-b_{11};y_1\right)-G\left(-b_5,-b_8;y_1\right)+G\left(-b_5,-b_9;y_1\right)+G\left(-b_5,-b_{10};y_1\right)\nonumber\\
   &-G\left(-b_5,-b_{11};y_1\right)-\log \left(\frac{1-x_{0,b}}{1+x_{0,b}}\right)
   \left(G\left(-b_4;y_1\right)-G\left(-b_5;y_1\right)\right)\Bigg]\nonumber\\
   &-i\pi \Bigg[\frac{x_{kl,-}+a_+ x_{kl,+}}{2\sqrt{a_+} \left(a_++1\right) \left(u_{kl,+}-u_+\right)}\left(G\left(-b_3;y_1\right)-G\left(-b_2;y_1\right)\right)\nonumber\\
   &+\frac{x_{kl,-}+a_-x_{kl,+}}{2 \sqrt{a_-} \left(a_-+1\right)\left(u_{kl,-}-u_+\right)}\left(G\left(-b_4;y_1\right)-G\left(-b_5;y_1\right)\right)\Bigg]\Bigg\}\times\frac{u_+-u_-}{\sqrt{A_{kl,b}}\sqrt{v_{kl}^{(+)}v_{kl}^{(-)}}}\,,
\end{align}
which contains $20$ GPLs. The expression for $\tilde{J}_1$ is:
\begin{align}
\tilde{J}_1=&\Bigg\{\frac{x_{kl,-}+a_+ x_{kl,+}}{2\sqrt{a_+} \left(a_++1\right)\left(u_{kl,+}-u_+\right)}\Bigg[2 G\left(-b_2,-b_1,-b_8;y_1\right)-2 G\left(-b_2,-b_1,-b_9;y_1\right)\nonumber\\
&-2G\left(-b_2,-b_1,-b_{10};y_1\right)+2G\left(-b_2,-b_1,-b_{11};y_1\right)-G\left(-b_2,-b_8,-b_8;y_1\right)\nonumber\\
&+G\left(-b_2,-b_8,-b_9;y_1\right)+G\left(-b_2,-b_8,-b_{10};y_1\right)-G\left(-b_2,-b_8,-b_{11};y_1\right)\nonumber\\
&-G\left(-b_2,-b_9,-b_8;y_1\right)+G\left(-b_2,-b_9,-b_9;y_1\right)+G\left(-b_2,-b_9,-b_{10};y_1\right)\nonumber\\
&-G\left(-b_2,-b_9,-b_{11};y_1\right)-G\left(-b_2,-b_{10},-b_8;y_1\right)+G\left(-b_2,-b_{10},-b_9;y_1\right)\nonumber\\
&+G\left(-b_2,-b_{10},-b_{10};y_1\right)-G\left(-b_2,-b_{10},-b_{11};y_1\right)-G\left(-b_2,-b_{11},-b_8;y_1\right)\nonumber\\
&+G\left(-b_2,-b_{11},-b_9;y_1\right)+G\left(-b_2,-b_{11},-b_{10};y_1\right)-G\left(-b_2,-b_{11},-b_{11};y_1\right)\nonumber\\
&-2G\left(-b_3,-b_1,-b_8;y_1\right)+2 G\left(-b_3,-b_1,-b_9;y_1\right)+2G\left(-b_3,-b_1,-b_{10};y_1\right)\nonumber\\
&-2G\left(-b_3,-b_1,-b_{11};y_1\right)+G\left(-b_3,-b_8,-b_8;y_1\right)-G\left(-b_3,-b_8,-b_9;y_1\right)\nonumber\\
&-G\left(-b_3,-b_8,-b_{10};y_1\right)+G\left(-b_3,-b_8,-b_{11};y_1\right)+G\left(-b_3,-b_9,-b_8;y_1\right)\nonumber\\
&-G\left(-b_3,-b_9,-b_9;y_1\right)-G\left(-b_3,-b_9,-b_{10};y_1\right)+G\left(-b_3,-b_9,-b_{11};y_1\right)\nonumber\\
&+G\left(-b_3,-b_{10},-b_8;y_1\right)-G\left(-b_3,-b_{10},-b_9;y_1\right)-G\left(-b_3,-b_{10},-b_{10};y_1\right)\nonumber\\
&+G\left(-b_3,-b_{10},-b_{11};y_1\right)+G\left(-b_3,-b_{11},-b_8;y_1\right)-G\left(-b_3,-b_{11},-b_9;y_1\right)\nonumber\\
&-G\left(-b_3,-b_{11},-b_{10};y_1\right)+G\left(-b_3,-b_{11},-b_{11};y_1\right)+\bigg(G\left(-b_2,-b_8;y_1\right)\nonumber\\
&+G\left(-b_2,-b_9;y_1\right)+G\left(-b_2,-b_{10};y_1\right)+G\left(-b_2,-b_{11};y_1\right)-G\left(-b_3,-b_8;y_1\right)\nonumber\\
&-G\left(-b_3,-b_9;y_1\right)-G\left(-b_3,-b_{10};y_1\right)-G\left(-b_3,-b_{11};y_1\right)\bigg)\log \left(\frac{1-x_{0,b}}{2}\right)\nonumber\\
&+\bigg(-G\left(-b_2,-b_8;y_1\right)-G\left(-b_2,-b_9;y_1\right)-G\left(-b_2,-b_{10};y_1\right)-G\left(-b_2,-b_{11};y_1\right)\nonumber\\
&+G\left(-b_3,-b_8;y_1\right)+G\left(-b_3,-b_9;y_1\right)+G\left(-b_3,-b_{10};y_1\right)+G\left(-b_3,-b_{11};y_1\right)\bigg) \log\left(\frac{1+x_{0,b}}{2}\right)\nonumber\\
&-2\left(G\left(-b_2,-b_1;y_1\right)-G\left(-b_3,-b_1;y_1\right)\right) \log\left(\frac{1-x_{0,b}}{1+x_{0,b}}\right)\nonumber\\
&+\frac{1}{2}\left(G\left(-b_2;y_1\right)-G\left(-b_3;y_1\right)\right) \left(\log^2\left(\frac{1-x_{0,b}}{2}\right)-\log^2\left(\frac{1+x_{0,b}}{2}\right)\right)\nonumber\\
&+\left(G\left(-b_2;y_1\right)-G\left(-b_3;y_1\right)\right) \left(\mathrm{Li}_2\left(\frac{1-x_{0,b}}{2}\right)-\mathrm{Li}_2\left(\frac{1+x_{0,b}}{2}\right)\right)\Bigg]\nonumber\\
&-\frac{x_{kl,-}+a_-x_{kl,+}}{2\sqrt{a_-} \left(a_-+1\right) \left(u_{kl,-}-u_+\right)}\Bigg[2 G\left(-b_4,-b_1,-b_8;y_1\right)-2 G\left(-b_4,-b_1,-b_9;y_1\right)\nonumber\\
&-2G\left(-b_4,-b_1,-b_{10};y_1\right)+2G\left(-b_4,-b_1,-b_{11};y_1\right)-G\left(-b_4,-b_8,-b_8;y_1\right)\nonumber\\
&+G\left(-b_4,-b_8,-b_9;y_1\right)+G\left(-b_4,-b_8,-b_{10};y_1\right)-G\left(-b_4,-b_8,-b_{11};y_1\right)\nonumber\\
&-G\left(-b_4,-b_9,-b_8;y_1\right)+G\left(-b_4,-b_9,-b_9;y_1\right)+G\left(-b_4,-b_9,-b_{10};y_1\right)\nonumber\\
&-G\left(-b_4,-b_9,-b_{11};y_1\right)-G\left(-b_4,-b_{10},-b_8;y_1\right)+G\left(-b_4,-b_{10},-b_9;y_1\right)\nonumber\\
&+G\left(-b_4,-b_{10},-b_{10};y_1\right)-G\left(-b_4,-b_{10},-b_{11};y_1\right)-G\left(-b_4,-b_{11},-b_8;y_1\right)\nonumber\\
&+G\left(-b_4,-b_{11},-b_9;y_1\right)+G\left(-b_4,-b_{11},-b_{10};y_1\right)-G\left(-b_4,-b_{11},-b_{11};y_1\right)\nonumber\\
&-2G\left(-b_5,-b_1,-b_8;y_1\right)+2 G\left(-b_5,-b_1,-b_9;y_1\right)+2G\left(-b_5,-b_1,-b_{10};y_1\right)\nonumber\\
&-2G\left(-b_5,-b_1,-b_{11};y_1\right)+G\left(-b_5,-b_8,-b_8;y_1\right)-G\left(-b_5,-b_8,-b_9;y_1\right)\nonumber\\
&-G\left(-b_5,-b_8,-b_{10};y_1\right)+G\left(-b_5,-b_8,-b_{11};y_1\right)+G\left(-b_5,-b_9,-b_8;y_1\right)\nonumber\\
&-G\left(-b_5,-b_9,-b_9;y_1\right)-G\left(-b_5,-b_9,-b_{10};y_1\right)+G\left(-b_5,-b_9,-b_{11};y_1\right)\nonumber\\
&+G\left(-b_5,-b_{10},-b_8;y_1\right)-G\left(-b_5,-b_{10},-b_9;y_1\right)-G\left(-b_5,-b_{10},-b_{10};y_1\right)\nonumber\\
&+G\left(-b_5,-b_{10},-b_{11};y_1\right)+G\left(-b_5,-b_{11},-b_8;y_1\right)-G\left(-b_5,-b_{11},-b_9;y_1\right)\nonumber\\
&-G\left(-b_5,-b_{11},-b_{10};y_1\right)+G\left(-b_5,-b_{11},-b_{11};y_1\right)+\bigg(G\left(-b_4,-b_8;y_1\right)\nonumber\\
&+G\left(-b_4,-b_9;y_1\right)+G\left(-b_4,-b_{10};y_1\right)+G\left(-b_4,-b_{11};y_1\right)-G\left(-b_5,-b_8;y_1\right)\nonumber\\
&-G\left(-b_5,-b_9;y_1\right)-G\left(-b_5,-b_{10};y_1\right)-G\left(-b_5,-b_{11};y_1\right)\bigg) \log \left(\frac{1-x_{0,b}}{2}\right)\nonumber\\
&+\bigg(-G\left(-b_4,-b_8;y_1\right)-G\left(-b_4,-b_9;y_1\right)-G\left(-b_4,-b_{10};y_1\right)-G\left(-b_4,-b_{11};y_1\right)\nonumber\\
&+G\left(-b_5,-b_8;y_1\right)+G\left(-b_5,-b_9;y_1\right)+G\left(-b_5,-b_{10};y_1\right)+G\left(-b_5,-b_{11};y_1\right)\bigg) \log\left(\frac{1+x_{0,b}}{2}\right)\nonumber\\
&-2\left(G\left(-b_4,-b_1;y_1\right)-G\left(-b_5,-b_1;y_1\right)\right) \log\left(\frac{1-x_{0,b}}{1+x_{0,b}}\right)\nonumber\\
&+\frac{1}{2}\left(G\left(-b_4;y_1\right)-G\left(-b_5;y_1\right)\right) \left(\log^2\left(\frac{1-x_{0,b}}{2}\right)-\log^2\left(\frac{1+x_{0,b}}{2}\right)\right)\nonumber\\
&+\left(G\left(-b_4;y_1\right)-G\left(-b_5;y_1\right)\right) \left(\mathrm{Li}_2\left(\frac{1-x_{0,b}}{2}\right)-\mathrm{Li}_2\left(\frac{1+x_{0,b}}{2}\right)\right)\Bigg]\nonumber\\
&+i\pi \Bigg[\frac{x_{kl,-}+a_+ x_{kl,+}}{2\sqrt{a_+} \left(a_++1\right)\left(u_{kl,+}-u_+\right)}\Bigg(2G\left(-b_2,-b_1;y_1\right)-G\left(-b_2,-b_8;y_1\right)-G\left(-b_2,-b_9;y_1\right)\nonumber\\
&-G\left(-b_2,-b_{10};y_1\right)-G\left(-b_2,-b_{11};y_1\right)-2G\left(-b_3,-b_1;y_1\right)+G\left(-b_3,-b_8;y_1\right)\nonumber\\
&+G\left(-b_3,-b_9;y_1\right)+G\left(-b_3,-b_{10};y_1\right)+G\left(-b_3,-b_{11};y_1\right)\nonumber\\
&+\left(G\left(-b_3;y_1\right)-G\left(-b_2;y_1\right)\right)\log\left(\frac{1-x_{0,b}^2}{4}\right)\Bigg)\nonumber\\
&-\frac{x_{kl,-}+a_-x_{kl,+}}{2\sqrt{a_-} \left(a_-+1\right) \left(u_{kl,-}-u_+\right)}\Bigg(2G\left(-b_4,-b_1;y_1\right)-G\left(-b_4,-b_8;y_1\right)-G\left(-b_4,-b_9;y_1\right)\nonumber\\
&-G\left(-b_4,-b_{10};y_1\right)-G\left(-b_4,-b_{11};y_1\right)-2G\left(-b_5,-b_1;y_1\right)+G\left(-b_5,-b_8;y_1\right)\nonumber\\
&+G\left(-b_5,-b_9;y_1\right)+G\left(-b_5,-b_{10};y_1\right)+G\left(-b_5,-b_{11};y_1\right)\nonumber\\
&+\left(G\left(-b_5;y_1\right)-G\left(-b_4;y_1\right)\right)\log\left(\frac{1-x_{0,b}^2}{4}\right)\Bigg)\Bigg]\Bigg\}\times\frac{u_+-u_-}{\sqrt{A_{kl,b}}\sqrt{v_{kl}^{(+)}v_{kl}^{(-)}}}\,,
\end{align}
which depends on 104 GPLs. The expressions for $\tilde{J}_i$ with $i\geq 2$ are too long to be written explicitly here, but can be found in the {\sc\small Mathematica} expressions in the ancillary file {\tt TwoMassiveEikonal\_FullAzimuth.wl}. There are $524$ and $2624$ GPLs in $\tilde{J}_2$ and $\tilde{J}_3$, respectively. The method presented here is not tailored for specific orders in $\epsilon$. Although higher-order terms in $\epsilon$ can certainly be computed without difficulty, we restrict ourselves to $\mathcal{O}(\epsilon^3)$ here.

Similar to the azimuthal-angle-averaged case, we compare our analytic expressions in terms of GPLs with the numerical evaluation of the one-fold integrals for $\tilde{J}_i$, using the {\sc\small Mathematica} function {\tt NIntegrate} at more than 4000 randomly sampled phase-space points. To evaluate the GPLs, we employ both {\tt handyG} and {\tt PolyLogTools}, leveraging their complementary features and compensating for their respective limitations. We find perfect agreement between our analytic results and the numerical integration. Furthermore, numerically performing the azimuthal-angle average reproduces our azimuthal-angle-averaged result. Finally, our azimuthal-angle-dependent one-loop soft function also agrees--up to $\mathcal{O}(\epsilon)$--with the revised result~\footnote{Z.~L. Liu, private correspondence} presented in ref.~\cite{Liu:2024hfa}, and matches those in refs.~\cite{Catani:2021cbl,Ju:2022wia} up to $\mathcal{O}(\epsilon^0)$.

\section{Summary\label{sec:summary}}

In this paper, we have presented an analytic calculation of the one-loop TMD soft function in impact-parameter space at higher orders in the dimensional regulator $\epsilon$. The resultant soft function (cf. eq.~\eqref{eq:TMD1Lsoftfuns}) is applicable to arbitrary scattering processes at hadron colliders only involving colored massive particles and a generic color-singlet system in the final state. 

An exception arises when the two initial-state partons have different Casimir factors, $C(\Ione)\neq C(\Itwo)$.~\footnote{Our results should also apply to quark-quark or antiquark-antiquark initiated processes, which can arise in BSM scenarios such as monotop production~\cite{Andrea:2011ws,Cacciapaglia:2018rqf}.} Even in these exceptional cases, our one-loop TMD soft function up to $\mathcal{O}(\epsilon^0)$ remains sufficient for obtaining the NLO soft function, which is adequate for NNLL resummation. However, due to rapidity divergences in the product of the two TMD beam functions, higher-order terms in $\alpha_1$ are required for an N$^k$LO-level ($k\geq 2$) soft function in (anti)quark-gluon initiated processes.

Our results, both the azimuthal-angle-averaged and the azimuthal-angle-dependent, are expressed in terms of multiple polylogarithms and are provided in a machine-readable form.  These analytic expressions can be further simplified using symbol techniques~\cite{Goncharov:2010jf,Duhr:2019tlz,Duhr:2011zq}, though we leave them in their current form without additional simplification efforts. Additionally, we present the bare soft function without performing UV (i.e., $\alpha_s$) renormalization or the subtraction of IR poles. The NLO soft function is already complete with our results up to $\mathcal{O}(\epsilon^0)$, where $\alpha_s$ renormalization does not contribute, and the $\msbar$ subtraction of IR poles is straightforward. The $\mathcal{O}(\epsilon^i)$ terms of our integrals contribute to a complete N$^k$LO calculation for $k\geq i+1$, for example through the nonabelian exponentiation theorem \cite{Gatheral:1983cz,Frenkel:1984pz},~\footnote{For instance, this behavior has been observed for the threshold soft function at NNLO in Laplace space~\cite{Ferroglia:2012uy,Wang:2018vgu}. However, the nonabelian exponentiation theorem applies only to the azimuthal-angle-dependent soft function in impact-parameter space. It does not hold in momentum space or for the azimuthal-angle-averaged case.} as well as through UV renormalization and IR subtraction. Nevertheless, these terms are not sufficient on their own, as they represent only one of the simplest components of a full N$^k$LO computation of the TMD soft function.

\begin{acknowledgments}
We are grateful to Ze Long Liu and Ding Yu Shao for helpful discussions regarding refs.~\cite{Liu:2024hfa} and \cite{Li:2013mia}, respectively. This work is supported by the ERC (grant 101041109 ``BOSON") and the French ANR (grant
ANR-20-CE31-0015, ``PrecisOnium"). Views and opinions expressed are however those of the authors only and do not necessarily reflect those of the European Union or the European Research Council Executive Agency. Neither the European Union nor the granting authority can be held responsible for them.
\end{acknowledgments}

\bibliographystyle{myutphys}
\bibliography{reference}

\end{document}